\documentclass[10pt,journal]{IEEEtran}

\usepackage{amssymb}
\usepackage{amsmath}
\usepackage{cite}
\usepackage{url}
\usepackage{xcolor}
\usepackage{cite,graphicx,amsmath,amssymb}
\usepackage{fancyhdr}
\usepackage{mdwmath}
\usepackage{mdwtab}
\usepackage{caption}
\usepackage{amsthm}
\usepackage{setspace}
\usepackage{hyperref}
\usepackage{algorithm}
\usepackage{algorithmic}
\usepackage{multirow}
\usepackage{makecell}
\usepackage{mathtools}
\usepackage{subcaption}
\usepackage{bm}
\usepackage{tikz}
\usepackage{xurl}
\usepackage{stfloats}
\usepackage{mathrsfs}

\usepackage{booktabs}
\usepackage{multirow}
\usepackage{siunitx}
\usepackage{balance}

\usepackage{booktabs}

\usepackage{multirow}

\hypersetup{colorlinks=true,
linkcolor=blue,
citecolor=blue,      
urlcolor=black,
}

\graphicspath{{./src/img/}}

\newtheorem{theorem}{Theorem}

\newtheorem{lemma}{Lemma}

\newtheorem{corollary}{Corollary}

\allowdisplaybreaks
\NewDocumentCommand{\multiubrace}{mmm}
 {% #1 = items, #2 = descriptions, #3 = separators
  \egreg_multiubrace:nnn {#1} {#2} {#3}
 }

\title{KDGen-BF: A Generative Site-Specific 

Multi-User Beamforming Approach}

\author{
        Ruihang Jiang, Zhaolin Wang,~\IEEEmembership{Member, IEEE}, and Yuanwei Liu,~\IEEEmembership{Fellow, IEEE}
\thanks{The authors are with the Department of Electrical and Computer Engineering, The University of Hong Kong, Hong Kong (e-mail: ruihang.jiang@connect.hku.hk, zhaolin.wang@hku.hk, yuanwei@hku.hk).}
}

\begin{document}

\bstctlcite{IEEEexample:BSTcontrol}

\maketitle

\begin{abstract}

% This paper proposes KDGen-BF, a knowledge-distillation and exponential-moving-average (KD-EMA) conditional diffusion transformer (DiT) framework for multi-user site-specific beamforming. 
% KDGen-BF infers joint multi-user beamformers from low-dimensional channel probing observations, i.e., reference signal received power (RSRP), without requiring full channel state information (CSI).
% To address ambiguity caused by limited probing observations and interference coupling among users, KDGen-BF formulates joint beamforming as a conditional generation problem and predicts continuous-domain multi-user beamformers from multi-user RSRP prompts.
% The framework combines a conditional DiT denoiser, teacher-student distillation training, and denoising diffusion implicit models (DDIM)-based multi-candidate generation.
% Experiments on multiple DeepMIMO scenarios demonstrate that: 1) at large probing budgets, KDGen-BF outperforms all non-exhaustive baselines and remains close to exhaustive search over the discrete Fourier transform (DFT) codebook; 2) at limited probing budgets, KDGen-BF outperforms all baselines; and 3) when RSRP observations are corrupted by noise, KDGen-BF maintains strong robustness and outperforms all compared baselines.

This paper proposes knowledge-distilled generative beamforming (KDGen-BF) framework for site-specific multi-user beamforming. KDGen-BF generates a multi-user beamforming weights from low-dimensional reference signal received power (RSRP) observations without acquiring instantaneous channel state information (CSI). To address the ambiguity caused by limited RSRP observations and interference coupling, KDGen-BF formulates multi-user beamforming as a conditional generation problem and directly outputs beamforming weights beyond a finite codebook. A diffusion transformer is trained through knowledge-distillation and exponential-moving-average (KD-EMA) guidance, and multi-candidate strategy is used for online deployment. Numerical results on multiple DeepMIMO scenarios demonstrate that: 1) under limited probing budgets, KDGen-BF outperforms all baselines; 2) with larger probing budgets, KDGen-BF achieves performance comparable to exhaustive search over the discrete Fourier transform (DFT) codebook and outperforms all other baselines; and 3) under noisy RSRP observations, KDGen-BF remains robust and outperforms all compared baselines.

\end{abstract}


\section{Introduction}
\IEEEPARstart{I}{N} multi-antenna wireless communications, beamforming is a cornerstone because it effectively utilizes spatial degrees of freedom~\cite{mietzner2009multiple}. The base station (BS) can steer the transmitted signal toward a specific user equipment (UE), leading to higher desired signal power. This directional transmission can compensate for the severe propagation and penetration losses in high frequency wireless systems. Beamforming is even more critical in multi-user downlink transmission, where users share the same time-frequency resource. The transmitted beams must be jointly designed to enhance desired signals while controlling inter-user interference.

Such joint beam design traditionally relies on accurate instantaneous channel state information (CSI) at the BS, since the channels of all users determine both desired signal enhancement and interference suppression~\cite{hassibi2003training}. Given CSI, conventional beamforming optimization methods, including weighted minimum mean square error (WMMSE)-based beamforming~\cite{shi2011iteratively}, zero-forcing-type precoding~\cite{wiesel2008zero}, and max--min signal-to-interference-plus-noise ratio (SINR) beamforming optimization~\cite{bjornson2014optimal}, can compute the beamforming weights in multi-user scenarios. However, acquiring accurate CSI in large array systems requires frequent pilot transmission and instantaneous channel estimation, which incurs substantial training overhead, baseband processing cost, and decision latency~\cite{shen2017high}. This overhead becomes more pronounced in multi-user systems, where the BS must acquire CSI for multiple users. Therefore, such CSI acquisition becomes a major bottleneck when beamforming decisions must be made online within each coherence block.

To avoid the requirement of instantaneous CSI, practical systems commonly adopt grid-of-beams (GoB)~\cite{giordani2019tutorial}, a codebook-based beamforming strategy that can be applied to both single-user and multi-user systems. In this paradigm, the BS transmits predefined probing beams and UEs report corresponding measurements, such as reference signal received power (RSRP). Nevertheless, GoB methods restrict the final beamforming solution to a finite set of predefined beams, leading to quantization loss~\cite{love2008overview}. Meanwhile, in the multi-user case, a more complicated tradeoff arises. A simple strategy is per-user greedy selection, where each UE selects its own probing beam with the strongest RSRP~\cite{giordani2019tutorial}. Although this approach exhibits low complexity, it ignores inter-user interference. An interference considered alternative is joint exhaustive search over all possible beam combinations in the codebook. However, the search space grows combinatorially with the number of users, resulting in high decision latency. Hierarchical search~\cite{alkhateeb2014channel,noh2020fast} reduces the sweeping overhead by using coarse-to-fine probing and keeping a small candidate beam set for each user. Nevertheless, it is still restricted to a discrete predefined beam grid. 

\subsection{Prior Works}

Conventional codebook-based beamforming reduces CSI acquisition through probing feedback, but it remains limited by predefined beam grids and, in multi-user systems, by the combinatorial search over joint beam combinations.
Site-specific beamforming (SSBF)~\cite{heng2024sitespecific} offers a different way to reduce online CSI acquisition and beam search overhead. The key idea of SSBF is to utilize the structure of a fixed propagation environment as side information. One representative direction of SSBF focuses on site-specific codebook learning, where probing beams are optimized from environmental data. In~\cite{zhang2020learning}, a neural network based framework was proposed to learn environment aware beam codebooks for millimeter-wave (mmWave) MIMO systems. In~\cite{wang2021sitespecific}, an online compressive beam codebook learning method was developed for mmWave vehicular communication by exploiting site-specific channel observations. In~\cite{heng2022learning}, site-specific probing beams were learned to reduce the beam alignment overhead, and the measured probing responses were used to predict the optimal narrow beam.
Unlike traditional discrete Fourier transform (DFT) codebooks, learned codebooks can better match the dominant propagation patterns of a given site and increase the informativeness of beam measurements. However, the final beamforming decision is still made over a finite set of candidate beams.

Another line of SSBF research moves beyond finite codebook selection and predicts continuous domain beamforming variables from incomplete observations. Discriminative neural networks can be used to map probing measurements to beamforming weights directly. 
In~\cite{heng2024gridfree}, a grid-free site-specific beam alignment framework was proposed, where learned probing beams and multilayer perceptron (MLP)-based beam synthesizers were jointly trained to directly synthesize transmit and receive beams from a continuous beam space.
In~\cite{li2025sitespecific}, site-specific beam learning was extended to full-duplex massive MIMO systems, where joint probing codebook design and beam synthesis were considered for transmit and receive beamforming.
These grid-free predictors enlarge the solution space. However, most of them are discriminative models that produce a single deterministic output and suffer from ambiguity. More specifically, similar RSRP responses may correspond to multiple plausible channels, and regression will lead to degraded beamforming performance.

Generative site-specific beamforming (GenSSBF)~\cite{wang2026generative} has recently been introduced to address the ambiguity in beam prediction from low-dimensional observations. Instead of producing a single deterministic beam from a low-dimensional RSRP observation, GenSSBF formulates beam prediction as conditional generation and outputs multiple plausible beam candidates. This is suitable when the same probing observation may correspond to multiple high-quality beamforming solutions. For example, in~\cite{zhou2026beam}, a U-Net-based Beam-Brainstorm framework was developed to generate multiple beamforming weights from RSRP.

\subsection{Motivation and Contributions}

However, existing GenSSBF studies mainly focus on single user beamforming and cannot be directly extended to the multi-user setting. In multi-user beamforming, a beam selected for one user can interfere with other users, so the beams cannot be generated independently. Moreover, there may be more than one optimal set of joint beamforming weights for the same multi-user channel, so forcing the model to learn only one output can be restrictive. These challenges motivate us to develop a joint conditional generation framework for online multi-user beamforming without CSI.

Therefore, we propose KDGen-BF, a conditional diffusion transformer framework that generates joint multi-user beamforming weights from RSRP observations. The proposed framework integrates a conditional diffusion transformer (DiT)~\cite{peebles2023scalable} beam generator, knowledge-distillation and exponential-moving-average (KD-EMA) training~\cite{tarvainen2017mean}, and denoising diffusion implicit model (DDIM)-based~\cite{song2021denoising} multi-candidate inference.
The main contributions are summarized as follows:
\begin{itemize}
    \item We propose KDGen-BF, a site-specific conditional generative framework for online multi-user beamforming without CSI.
    Given multi-user RSRP observations, KDGen-BF directly generates joint beamforming weights rather than selecting beams from a finite codebook.

    \item We design a second-order-cone (SOC)-guided~\cite{gershman2010convex} KD-EMA training strategy to solve the ambiguity of low-dimensional RSRP observations and the multi-solution nature of joint beamforming.
    The SOC beamforming weights computed with full CSI are used only as offline teacher supervision, while KD-EMA training transfers this supervision to the DiT beam generator and improves robustness through masked learning.

    \item We develop a deployable DDIM-based multi-candidate inference strategy with feedback based candidate selection.
    The proposed inference strategy generates multiple candidate beamformers online without instantaneous CSI acquisition at the BS, and selects the final beamforming matrix using lightweight UE feedback.
    
    \item We conduct experiments on multiple DeepMIMO scenarios to evaluate the proposed framework.
    The results demonstrate that KDGen-BF achieves max--min fairness gains over codebook-based, learned-codebook, deterministic, and single-user-direction baselines, especially under limited probing budgets and noisy RSRP observations.
\end{itemize}

\subsection{Organization and Notations}

The remainder of this paper is organized as follows.
Section~\ref{sec:problem} presents the system model and problem formulation.
Section~\ref{sec:algorithm} introduces the proposed KDGen-BF framework.
Section~\ref{sec:baseline_compare} describes the baseline methods and complexity comparison.
Section~\ref{sec:results} presents the numerical results, and Section~\ref{sec:conclusion} concludes the paper.

\emph{Notations:}
Scalars, vectors, and matrices are denoted by italic, bold lowercase, and bold uppercase letters, respectively.
For a matrix $\mathbf A$, $\mathbf A^T$, $\mathbf A^H$, and $\mathrm{Tr}(\mathbf A)$ denote its transpose, Hermitian transpose, and trace, respectively.
The notations $\mathbf A_{i,:}$ and $\mathbf A_{:,j}$ denote the $i$-th row and the $j$-th column of $\mathbf A$, respectively.
The Euclidean and Frobenius norms are denoted by $\|\cdot\|_2$ and $\|\cdot\|_F$, respectively.
The expectation operator and computational complexity order are denoted by $\mathbb E[\cdot]$ and $\mathcal O(\cdot)$, respectively.
Other symbols are defined when they first appear.

\section{System Model and Problem Formulation}
\label{sec:problem}

\begin{figure*}[t]
    \centering
    \includegraphics[width=0.9\textwidth]{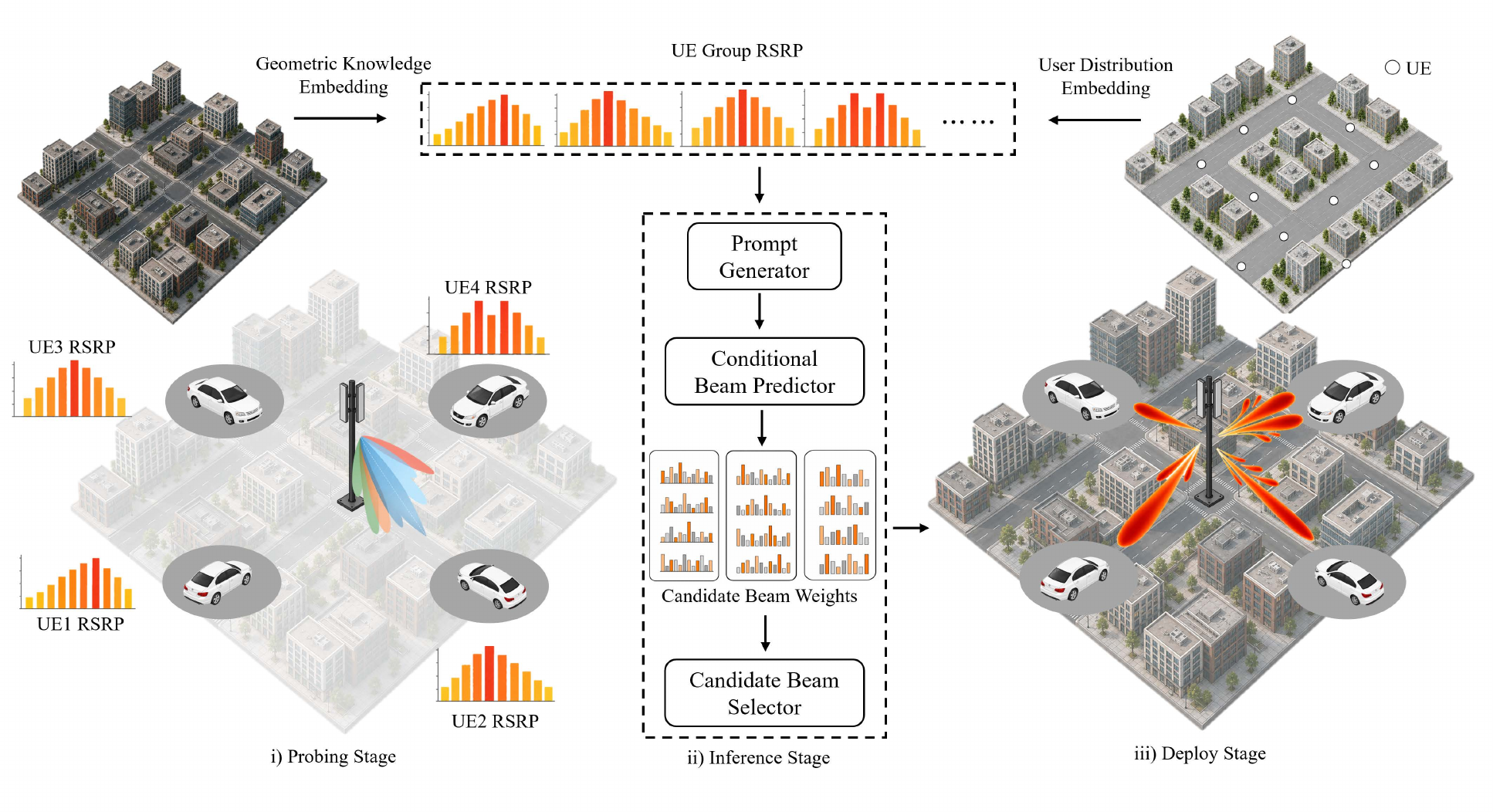}
    \caption{Three-stage site-specific multi-user downlink beamforming framework. 
    The BS first collects channel probing RSRP feedback, then generates a joint beamforming weights from the multi-user prompt without requiring instantaneous CSI at the BS, and finally deploys the generated beams for simultaneous downlink transmission.}
    \label{fig:scenario}
\end{figure*}

\subsection{System Model}
\label{sec:system_model}

As illustrated in Fig.~\ref{fig:scenario}, we consider a single cell multi-user downlink MISO system, where a BS equipped with an $M_t$-element uniform linear array (ULA) serves $K$ single-antenna UEs indexed by $\mathcal{K}=\{1,\ldots,K\}$. A narrowband quasi-static block-fading model is considered, under which the channel remains constant within each coherence block. In this work, we assume the whole beamforming procedure is performed within a single coherence block, such that the channel is treated as fixed during the online process.
% The single-antenna user setting is adopted to isolate the core problem of interest, namely, BS-side transmit beam prediction from probing-induced observations, without introducing additional receive-side combining variables or user-side transceiver design coupling.

The downlink channel from the BS to user $k$ is given by $\mathbf{h}_k\in\mathbb{C}^{M_t\times 1}$ and modeled as
\begin{equation}
\mathbf{h}_k=\sum_{\ell=1}^{L_k}\alpha_{k,\ell}\mathbf{a}\!\left(\phi_{k,\ell}^{D}\right),
\end{equation}
where $L_k$ denotes the number of propagation paths associated with user $k$, $\alpha_{k,\ell}\in\mathbb{C}$ is the complex gain of the $\ell$-th path, and $\phi_{k,\ell}^{D}$ is its azimuth angle of departure (AoD) at the BS.
For a ULA with inter-element spacing $d$, the array steering vector under the narrowband far field assumption is given by
\begin{equation}
\mathbf{a}(\phi)=\frac{1}{\sqrt{M_t}}
\left[
1,\,
e^{j2\pi\frac{d}{\lambda}\sin(\phi)},\,
\ldots,\,
e^{j2\pi\frac{(M_t-1)d}{\lambda}\sin(\phi)}
\right]^T,
\end{equation}
where $\lambda$ denotes the carrier wavelength. 
% The normalization ensures that $\|\mathbf{a}(\phi)\|_2^2=1$. 
% The aggregate channel matrix is defined as $\mathbf{H}=[\mathbf{h}_1,\ldots,\mathbf{h}_K]^H\in\mathbb{C}^{K\times M_t}$.

% Regarding the transceiver architecture, the number of simultaneously multiplexed data streams is limited by the number of RF chains. Since this work considers one independent data stream for each user over the same time--frequency resource block, the transceiver must satisfy $K\leq N_{\mathrm{RF}}\leq M_t$. Under this formulation, pure analog beamforming, which typically employs a single RF chain, is not suitable for general multi-user spatial multiplexing. Hybrid beamforming corresponds to $1<N_{\mathrm{RF}}<M_t$, whereas fully-digital beamforming uses one RF chain per antenna, i.e., $N_{\mathrm{RF}}=M_t$. In this work, we focus on the fully-digital setting in order to characterize the beamforming performance Upper Bound and isolate the core algorithmic behavior from hardware-induced constraints.

% The beamforming matrix is defined as $\mathbf{W}=[\mathbf{w}_1,\ldots,\mathbf{w}_K]\in\mathbb{C}^{M_t\times K}$, and 

Let $s_k$ denote the transmit symbol intended for user $k$, satisfying $\mathbb{E}[|s_k|^2]=1$ and $\mathbb{E}[s_k s_i^*]=0$ for $i\neq k$, and $\mathbf{w}_k\in\mathbb{C}^{M_t\times 1}$ denotes the corresponding beamforming vector.
The transmitted signal at the BS is given by
\begin{equation}
\mathbf{x}=\sum_{k=1}^{K}\mathbf{w}_k s_k.
\end{equation}

Based on the above model, the received signal at user $k$ can be expressed as
\begin{equation}
y_k
= \mathbf{h}_k^H\mathbf{x}+n_k
= \mathbf{h}_k^H\mathbf{w}_k s_k
+ \sum_{i\neq k}\mathbf{h}_k^H\mathbf{w}_i s_i + n_k,
\end{equation}
where $n_k\sim\mathcal{CN}(0,\sigma^2)$ denotes the additive white Gaussian noise (AWGN) at user $k$. Accordingly, the signal-to-interference-plus-noise ratio (SINR) of user $k$ is given by
\begin{equation}
\gamma_k(\mathbf{W})=
\frac{|\mathbf{h}_k^H\mathbf{w}_k|^2}
{\sum_{i\neq k}|\mathbf{h}_k^H\mathbf{w}_i|^2+\sigma^2}.
\label{eq:sinr_define}
\end{equation}

Based on the SINR in \eqref{eq:sinr_define}, the achievable rate of user $k$ is given by
\begin{equation}
R_k(\mathbf{W})=\log_2\!\left(1+\gamma_k(\mathbf{W})\right).
\end{equation}
where $\mathbf{W}=[\mathbf{w}_1,\ldots,\mathbf{w}_K]\in\mathbb{C}^{M_t\times K}$ is the overall multi-user beamforming matrix.
Since $\gamma_k(\mathbf{W})$ contains inter-user interference terms, the achievable rate of each user depends on the entire beamforming matrix $\mathbf{W}$ rather than only on its own beamforming vector.

\subsection{Problem Formulation}

In this paper, we consider the following max--min beamforming design problem:
\begin{equation}
\begin{aligned}
\max_{\mathbf{W}} \quad & \min_{k\in\mathcal{K}} R_k(\mathbf{W})\\
\text{s.t.}\quad
& \sum_{k=1}^{K}\|\mathbf{w}_k\|_2^2\le P_{\mathrm{tot}},\\
& \mathbf{W}\in\mathcal{S},
\end{aligned}
\label{eq:maxmin_problem}
\end{equation}
where $P_{\mathrm{tot}}$ denotes the BS transmit power budget, and $\mathcal{S}$ denotes the feasible beamforming set determined by the adopted transceiver architecture. 
% In the fully-digital setting considered in this work, no additional structural constraint is imposed on $\mathbf{W}$ beyond the transmit power constraint, i.e., $\mathcal{S}=\mathbb{C}^{M_t\times K}$. Although the subsequent development focuses on the fully-digital case, the same formulation can in principle be extended to hybrid beamforming by redefining the feasible set and the target beamforming solutions under the associated hardware constraints.

The optimization problem in \eqref{eq:maxmin_problem} is challenging due to the coupled interference structure across users and the resulting non-convexity of the objective. In particular, each user rate depends on the entire beamforming matrix through the inter-user interference terms, which prevents the problem from being decomposed into independent per-user subproblems. If CSI were available at the BS, \eqref{eq:maxmin_problem} could in principle be addressed directly using  conventional beamforming optimization methods~\cite{bjornson2014optimal}. The main challenge considered in this work is that such CSI is unavailable during online deployment, and the beamforming decision must instead be made from limited channel probing observations. This motivates a site-specific learning-based reformulation that predicts beamforming matrices directly from low-dimensional channel probing observations.

% \begin{remark}
% This work focuses on the interference-relevant regime, where inter-user interference is non-negligible and joint beam coordination is essential for fairness-aware multi-user transmission.
% In this regime, each user's rate depends not only on its own beamforming vector but also on the beams assigned to the other users, which makes independent per-user beam selection insufficient.
% When thermal noise dominates, the system approaches the low-signal-to-noise-ratio (SNR) noise-limited regime, where the optimal transmit beamforming direction tends to maximum-ratio transmission (MRT), and the problem becomes closer to independent single-user beam alignment~\cite{bjornson2014optimal}.
% \end{remark}

\subsection{Site-Specific Learning Reformulation}

We next instantiate this reformulation by first specifying the channel probing observations and the site-specific structure used for learning. In practical beamforming problem, the probing observations are obtained through beam probing: the BS transmits a set of predefined probing beams, and each UE reports the corresponding RSRP values. These measurements provide only a low-dimensional response, rather than the full complex channel vector, but they indicate how each UE responds to the probed directions.

The site-specific assumption provides the missing structure needed to use these coarse observations for beamforming. For a fixed deployment, the BS location, dominant scatterers, blockage patterns, and long-term user distribution remain relatively stable. As a result, RSRP patterns observed in the same site are statistically related to effective beamforming decisions. Site-specific learning exploits this relation to infer joint beamforming weights from RSRP observations, thereby avoiding online CSI acquisition and exhaustive beam search.

% Motivated by the limitations of finite-codebook search and existing single-user or deterministic GenSSBF methods, we reformulate CSI-free multi-user beamforming as conditional generation from probing observations to joint beamforming weights.
% This reformulation avoids explicit combinatorial tuple search, removes the restriction to a predefined beam grid, and allows the model to represent multiple high-utility beamforming solutions under limited RSRP observations.
% Here, CSI-free means that the BS does not require instantaneous channel vectors during online beam generation and selection; low-dimensional UE feedback, such as probing RSRP values and candidate-level utility reports, is still used.

Specifically, let $\mathbf{p}_k=[p_{k,1},\ldots,p_{k,N_p}]^T\in\mathbb{R}^{N_p\times 1}$ denote the RSRP vector of user $k$, where $N_p$ is the number of DFT beams used for probing, and $p_{k,n}$ represents the measured RSRP corresponding to the $n$-th probing beam. Let $\mathcal{P}=\{\mathbf{p}_k\}_{k=1}^{K}$ denote the resulting multi-user RSRP observation set, and define $\mathcal{H}=\{\mathbf{h}_k\}_{k=1}^{K}$ as the multi-user channel set. To characterize the site-specific setting, we conceptually model the probing observation process as
\begin{equation}
\mathcal{P}=\mathcal{M}_{\mathcal{E}}(\mathcal{H}),
\end{equation}
where $\mathcal{E}$ denotes the site related propagation environment, including the BS deployment, the long-term user spatial distribution within the site, dominant scatterers or blockages, and other environment related propagation characteristics; $\mathcal{M}_{\mathcal{E}}(\cdot)$ is the corresponding channel probing operator.

Based on $\mathcal{P}$, we further construct a multi-user prompt representation $\mathbf{O}$, which serves as the input to the subsequent beam prediction framework. In the site-specific setting considered in this work, $\mathcal{E}$ is fixed for a given deployment scenario. Therefore, $\mathcal{E}$ is neither explicitly estimated nor provided to the network during inference; instead, its effect is implicitly embedded in $\mathbf{O}$ through the channel probing RSRP observations. Based on the prompt representation $\mathbf{O}$, the joint beamforming matrix is predicted as
\begin{equation}
\hat{\mathbf{W}} = f_{\theta}(\mathbf{O}), \qquad
\hat{\mathbf{W}} \in \mathbb{C}^{M_t\times K},
\end{equation}
where $f_\theta(\cdot)$ is defined as the learnable mapping from the multi-user prompt to the joint beamforming matrix. A decoding and normalization module maps the generated representation to beamforming weights and enforces the total transmit power constraint and the feasibility constraints imposed by the adopted transmitter architecture.

To learn the relation from the multi-user prompt $\mathbf{O}$ to the joint beamforming weights $\hat{\mathbf{W}}$, training samples are collected from the considered site as paired prompt and channel samples $(\mathbf{O},\mathcal{H})$. Let $\gamma_k(\hat{\mathbf{W}};\mathcal{H})$ denote the SINR of user $k$ evaluated under channel set $\mathcal{H}$. The desired task objective is to learn a model that maximizes the expected worst user rate:
\begin{equation}
\max_{\theta}\;
\mathbb{E}_{(\mathbf{O},\mathcal{H})\sim p_{\mathrm{data}}}
\left[
\min_{k\in\mathcal{K}}
\log_2\!\left(1+\gamma_k(\hat{\mathbf{W}};\mathcal{H})\right)
\right].
\label{eq:task_objective}
\end{equation}

The objective in \eqref{eq:task_objective} evaluates the generated beamforming weights using the channel set $\mathcal{H}$. During online inference, however, the BS only receives the low-dimensional RSRP observation set $\mathcal{P}$ and constructs the prompt $\mathbf{O}$ from it; the instantaneous channel set $\mathcal{H}$ is not available at the BS. Thus, the online task is to generate $\hat{\mathbf{W}}$ from $\mathcal{P}$, or equivalently from $\mathbf{O}$, without observing $\mathcal{H}$. This task is difficult because limited RSRP observations may correspond to different channel conditions, and the same channel condition may admit multiple optimal sets of joint beamforming weights. To address this challenge, we propose KDGen-BF in the following section.
\begin{figure*}[!t]
    \centering
    \includegraphics[width=0.9\textwidth]{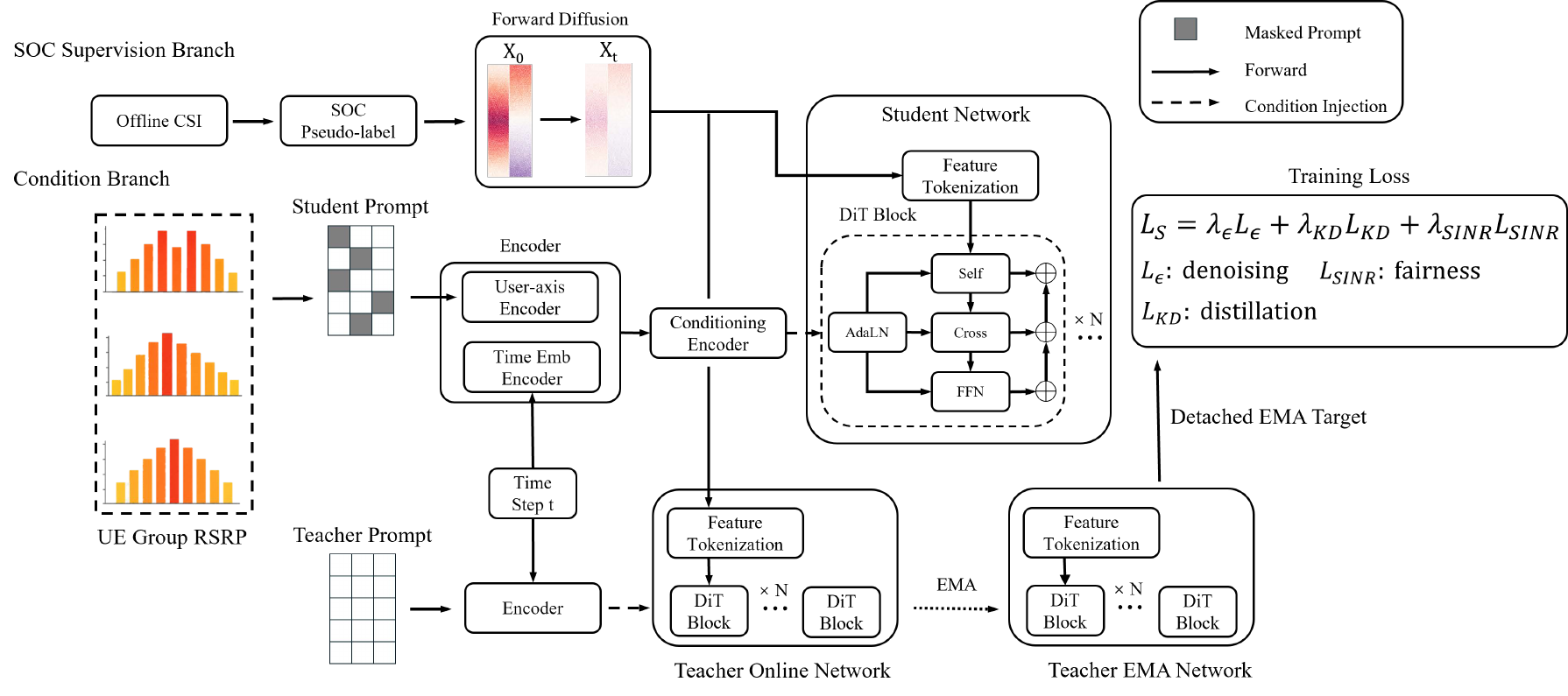}
    \caption{Training framework of KDGen-BF. Offline SOC solutions provide full CSI pseudo-labels for diffusion training, while the RSRP prompt and diffusion timestep condition the denoiser. The student is trained with masked prompts and guided by an EMA teacher through denoising, distillation, and fairness-oriented losses.}
    \label{fig:model_training}
\end{figure*}

\section{Proposed KDGen-BF for Site-Specific Multi-User Beamforming}\label{sec:algorithm}

\subsection{Framework Overview}
\label{sec:framework_overview}

As shown in Fig.~\ref{fig:model_training}, we propose a site-specific conditional generative framework for solving the max--min fair multi-user beamforming problem in \eqref{eq:maxmin_problem}. Given the RSRP observations reported by all users in a group, the framework predicts a joint beamforming matrix $\hat{\mathbf{W}}\in\mathbb{C}^{M_t\times K}$ without requiring instantaneous CSI at the BS during online inference.

The framework consists of three components. \emph{First}, the RSRP observations reported by all users are converted into a multi-user prompt for conditional beam generation. \emph{Second}, a conditional diffusion transformer (DiT) denoiser generates multi-user beamforming weights by using timestep information and multi-user dependencies. \emph{Third}, SOC-based pseudo-labels and EMA-stabilized teacher-student offline training, while DDIM sampling with multi-candidate generation enables online inference.
We first describe how the RSRP observations are converted into the multi-user prompt used by the conditional denoiser.

\subsection{Prompt Construction and Conditioning}
\label{sec:io_cond}

\subsubsection{Prompt construction}
We construct the model prompt from low-dimensional channel probing RSRP observations.
Let $\mathcal{C}_{\mathrm{p}}=\{\mathbf{f}_1,\ldots,\mathbf{f}_{N_p}\}\subseteq\mathcal{C}$ denote the probed DFT beam subset.
For each user $k\in\mathcal{K}$, let $\mathbf{p}_k=[p_{k,1},\ldots,p_{k,N_p}]^T\in\mathbb{R}^{N_p\times 1}$ represent the probing RSRP vector over $\mathcal{C}_{\mathrm{p}}$.

From $\mathbf{p}_k$, we extract a standardized probing feature and a peak referenced feature.
The standardized feature $\mathbf{z}_k\in\mathbb{R}^{N_p\times 1}$ is obtained by normalizing the entries of $\mathbf{p}_k$ using the mean and standard deviation computed within the same user's RSRP vector. This normalization emphasizes relative response variations and reduces scale differences across users.
The peak referenced feature $\mathbf{r}_k\in\mathbb{R}^{N_p\times 1}$ is obtained by measuring each entry relative to the strongest probed beam. This feature preserves the relative gap between the dominant probing direction and the other probing directions.

By concatenating these two feature views, the user prompt is formed as
$\mathbf{o}_k=[\mathbf{z}_k^T,\mathbf{r}_k^T]^T\in\mathbb{R}^{2N_p\times 1}$.
Stacking all user prompts gives the multi-user prompt matrix
$\mathbf{O}=[\mathbf{o}_1^T,\ldots,\mathbf{o}_K^T]^T\in\mathbb{R}^{K\times 2N_p}$.
Together, $\mathbf{z}_k$ and $\mathbf{r}_k$ provide a scale robust representation that preserves both relative RSRP variations and the contrast to the strongest probed beam.

\subsubsection{Prompt conditioning}
To condition the model on the group prompt, $\mathbf{O}$ is encoded into one context token for each user. Throughout this paper, the term \textit{token} refers to a vector representation processed by Transformer attention layers.
Since the fairness oriented beamforming decision depends on multi-user coupling rather than independent per-user preferences, the conditioning representation should capture not only per-user prompt features but also their cross-user dependency structure.
To this end, we employ a prompt encoder that applies self-attention across users and maps the multi-user prompt to
\begin{equation}
\mathbf{C}_u=E_u(\mathbf{O})\in\mathbb{R}^{K\times d_c},
\end{equation}
where $d_c$ denotes the context token dimension.
The resulting matrix $\mathbf{C}_u$ contains one context token for each user, while self-attention enables each token to incorporate dependency information from the other users in the same group.

The diffusion denoiser is conditioned not only on the multi-user prompt but also on the diffusion timestep.
At denoising step $t$, let $\mathbf{e}_t\in\mathbb{R}^{d\times 1}$ denote the timestep embedding, and $\mathbf{c}_t\in\mathbb{R}^{d_c\times 1}$ represents the corresponding time token obtained from $\mathbf{e}_t$ through a learnable projection.
We then form the cross-attention context as
\begin{equation}
\mathbf{C}
=
\begin{bmatrix}
\mathbf{c}_t^T\\
\mathbf{C}_u
\end{bmatrix}
\in\mathbb{R}^{(K+1)\times d_c}.
\label{eq:C_definition}
\end{equation}
This design keeps the conditioning tokens separate from the main denoising sequence and allows the subsequent diffusion model to inject timestep and prompt information in a structured way.

In addition to the token conditioning context prepared for subsequent cross-attention, we further construct a global conditioning signal for layer modulation.
Specifically, let the $k$-th user token be denoted by $\mathbf{c}_k^{(u)}\in\mathbb{R}^{d_c\times 1}$, where the $k$-th row of $\mathbf{C}_u$ is given by $(\mathbf{c}_k^{(u)})^T$.
Then, the global conditioning vector is constructed as
\begin{equation}
\mathbf{u}
=
\mathbf{e}_t
+
\mathrm{MLP}\!\left(
\frac{1}{K}\sum_{k=1}^{K}\mathbf{c}_k^{(u)}
\right)
\in\mathbb{R}^{d\times 1},
\label{eq:u_definition}
\end{equation}
where the MLP maps the average of the $K$ user context tokens from $\mathbb{R}^{d_c\times 1}$ to $\mathbb{R}^{d\times 1}$.
This global conditioning summarizes the overall multi-user context and provides a shared modulation signal across layers.

\subsection{Conditional DiT Denoiser}
\label{sec:dit_denoiser}

With the cross-attention context and global conditioning signal defined in Section~\ref{sec:io_cond}, we now describe the diffusion model that operates on noisy beam representations and reconstructs the target multi-user beamforming solution.

\subsubsection{Beamforming target representation}
Rather than directly predicting the complex beamforming matrix, we first convert the target beamforming solution into a real-valued DFT domain representation, denoted by $\mathbf{X}_0$. In the diffusion model, $\mathbf{X}_0$ is treated as the target sample before noise perturbation.
Specifically, define $\mathbf{X}_0\in\mathbb{R}^{2K\times M_t}$ as the target beam representation, where the first $K$ rows correspond to the real parts of the user beams and the remaining $K$ rows correspond to the imaginary parts.
For user $k\in\mathcal{K}$, the associated complex DFT domain beam is reconstructed as
\begin{equation}
\tilde{\mathbf{w}}_k^{\mathrm{dft}}
=
\left(\mathbf{X}_{0,k,:}
+
j\,\mathbf{X}_{0,K+k,:}\right)^T
\in\mathbb{C}^{M_t\times 1}.
\end{equation}

This representation is adopted for two reasons.
\emph{First}, the probing observations are collected from DFT beam measurements, so the conditioning prompt and the beam output are naturally aligned in the same DFT domain coordinate system.
\emph{Second}, beam patterns are typically more structured in the angular and spectral domain than in the spatial domain, which is beneficial for conditional generation and denoising.

The corresponding spatial domain beam is then obtained by inverse discrete Fourier transform as
$
\tilde{\mathbf{w}}_k
=
\mathrm{IFFT}\!\left(\tilde{\mathbf{w}}_k^{\mathrm{dft}}\right)
\in\mathbb{C}^{M_t\times 1}.
$
To satisfy the total transmit-power constraint, we further apply global normalization across all users:
\begin{equation}
\mathbf{w}_k
=
\sqrt{\frac{P_{\mathrm{tot}}}{\sum_{i=1}^{K}\|\tilde{\mathbf{w}}_i\|_2^2}}
\,\tilde{\mathbf{w}}_k,\qquad k\in\mathcal{K}.
\end{equation}
By concatenating the resulting user beams, we obtain the final beamforming matrix
$
\hat{\mathbf{W}}
=
[\mathbf{w}_1,\ldots,\mathbf{w}_K]
\in\mathbb{C}^{M_t\times K}.
$

\subsubsection{Forward diffusion process}
To train the denoiser, the target beam representation $\mathbf{X}_0$ is progressively corrupted through the standard forward diffusion process~\cite{ho2020denoising}.
At diffusion step $t$, the noisy sample $\mathbf{X}_t$ is drawn from
\begin{equation}
q(\mathbf{X}_t\mid \mathbf{X}_0)
=
\mathcal{N}\!\left(
\sqrt{\bar{\alpha}_t}\,\mathbf{X}_0,\,
(1-\bar{\alpha}_t)\mathbf{I}
\right),
\label{eq:forward_diffusion_distribution}
\end{equation}
The above representation can be equivalently written as
\begin{equation}
\mathbf{X}_t
=
\sqrt{\bar{\alpha}_t}\,\mathbf{X}_0
+
\sqrt{1-\bar{\alpha}_t}\,\boldsymbol{\epsilon},
\qquad
\boldsymbol{\epsilon}\sim\mathcal{N}(\mathbf{0},\mathbf{I}),
\label{eq:forward_diffusion_reparam}
\end{equation}
where $\boldsymbol{\epsilon}$ denotes the Gaussian noise injected into the target beam representation at diffusion step $t$.
Here, $\bar{\alpha}_t=\prod_{s=1}^{t}(1-\beta_s)$ denotes the cumulative noise coefficient, where $\{\beta_s\}_{s=1}^{T}$ is the variance schedule over the diffusion horizon.
Under the forward diffusion process defined above, the denoising task is to predict the injected noise from $\mathbf{X}_t$ under the conditioning representation defined in Section~\ref{sec:io_cond}.

\subsubsection{One-dimensional DiT architecture}
To perform conditional denoising, we adopt a one-dimensional DiT backbone operating on the noisy beam representation. Since $\mathbf{X}_t\in\mathbb{R}^{2K\times M_t}$ is represented in the DFT domain, its $M_t$ columns correspond to DFT beam indices. We therefore treat the DFT beam dimension as a one-dimensional sequence with length $M_t$, which is more suitable than a two-dimensional token representation designed for image inputs.
Let $d$ denote the hidden width of the DiT backbone.
Given $\mathbf{X}_t\in\mathbb{R}^{2K\times M_t}$, the input token sequence is initialized as
\begin{equation}
\mathbf{H}^{(0)}
=
\mathrm{Proj}_{\mathrm{in}}(\mathbf{X}_t^T)
+
\mathbf{P}_{\mathrm{pos}}
+
\mathbf{P}_{\mathrm{stem}}(\mathbf{X}_t^T),
\end{equation}
where $\mathrm{Proj}_{\mathrm{in}}:\mathbb{R}^{M_t\times 2K}\rightarrow\mathbb{R}^{M_t\times d}$ maps the noisy beam representation into the hidden token space, $\mathbf{P}_{\mathrm{pos}}\in\mathbb{R}^{M_t\times d}$ denotes the positional embedding, and $\mathbf{P}_{\mathrm{stem}}(\mathbf{X}_t^T)\in\mathbb{R}^{M_t\times d}$ provides shallow input related features.

The cross-attention context $\mathbf{C}$ and the global conditioning vector $\mathbf{u}$ are defined in \eqref{eq:C_definition} and \eqref{eq:u_definition}, respectively. Let $\ell\in\{0,\ldots,L-1\}$ represent the block index, where $L$ is the number of stacked DiT blocks.
Each DiT block then applies the following residual updates:
\begin{subequations}\label{eq:dit_block}
\begin{align}
\mathbf{H}_{\mathrm{sa}}^{(\ell)}
&=
\mathbf{H}^{(\ell)}
+
\mathrm{SA}\!\big(\mathrm{AdaLN}(\mathbf{H}^{(\ell)};\mathbf{u})\big),\\
\mathbf{H}_{\mathrm{ca}}^{(\ell)}
&=
\mathbf{H}_{\mathrm{sa}}^{(\ell)}
+
\mathrm{CA}\!\big(\mathrm{AdaLN}(\mathbf{H}_{\mathrm{sa}}^{(\ell)};\mathbf{u}),\mathbf{C}\big),\\
\mathbf{H}^{(\ell+1)}
&=
\mathbf{H}_{\mathrm{ca}}^{(\ell)}
+
\mathrm{FFN}\!\big(\mathrm{AdaLN}(\mathbf{H}_{\mathrm{ca}}^{(\ell)};\mathbf{u})\big).
\end{align}
\end{subequations}
Here, SA, CA, and FFN are self-attention, cross-attention, and the position feed-forward network, respectively.
In particular, self-attention captures structural correlation within the noisy beam sequence, cross-attention injects timestep and prompt context through $\mathbf{C}$, and AdaLN uses the global conditioning vector $\mathbf{u}$ to provide shared layer modulation.
This combination enables joint denoising of multi-user beam representations while preserving the structured one-dimensional nature of the beam sequence.
After the final block, the hidden sequence $\mathbf{H}^{(L)}$ is projected to the predicted noise tensor $\hat{\boldsymbol{\epsilon}}\in\mathbb{R}^{2K\times M_t}$ through an output head.

\subsubsection{Denoising objective and target beam reconstruction}
The predicted noise tensor $\hat{\boldsymbol{\epsilon}}$ produced by the DiT backbone is trained to match the injected Gaussian noise $\boldsymbol{\epsilon}$.
Accordingly, the base denoising objective is given by
\begin{equation}
L_{\mathrm{diff}}
=
\mathbb{E}_{\mathbf{X}_0,t,\boldsymbol{\epsilon}}
\!\left[
\left\|
\boldsymbol{\epsilon}
-
\hat{\boldsymbol{\epsilon}}
\right\|_F^2
\right].
\end{equation}
This objective trains the model to recover the target beam representation by predicting the additive noise from the corrupted sample under prompt conditioning.
From the predicted noise, the target estimate is reconstructed as
\begin{equation}
\label{eq:x0_reconstruction}
\hat{\mathbf{X}}_0
=
\frac{\mathbf{X}_t-\sqrt{1-\bar{\alpha}_t}\,\hat{\boldsymbol{\epsilon}}}{\sqrt{\bar{\alpha}_t}}.
\end{equation}
Substituting the forward diffusion relation of $\mathbf{X}_t$ into the above expression yields the reconstruction error
\begin{equation}
\label{eq:x0_error_relation}
\hat{\mathbf{X}}_0-\mathbf{X}_0
=
\sqrt{\frac{1-\bar{\alpha}_t}{\bar{\alpha}_t}}
\left(
\boldsymbol{\epsilon}
-
\hat{\boldsymbol{\epsilon}}
\right).
\end{equation}
Therefore, the reconstruction fidelity of the target beam representation is directly controlled by the noise prediction accuracy, with a step related amplification factor determined by $\bar{\alpha}_t$.
In particular, errors at high noise steps are more strongly amplified in the reconstructed target estimate.
The reconstructed $\hat{\mathbf{X}}_0$ is then mapped to the final beamforming matrix $\hat{\mathbf{W}}$ through the DFT domain decoding, inverse Fourier transform, and total-power normalization.

\subsection{SOC Teacher Construction and KD-EMA Training}
\label{sec:soc_kd}

\subsubsection{SOC-based pseudo-label construction}
To provide high-quality supervision for the diffusion beamformer, we construct pseudo-labels by solving the max--min beamforming problem using offline CSI.
For each user group with channel set $\mathbf{H}=\{\mathbf{h}_k\}_{k=1}^{K}$, we first solve the following max--min SINR problem under the total transmit-power constraint:
\begin{equation}
\begin{aligned}
\max_{\mathbf{W},\,\tau}\quad & \tau \\
\mathrm{s.t.}\quad
& \gamma_k(\mathbf{W}) \ge \tau,\qquad \forall k\in\mathcal{K},\\
& \sum_{k=1}^{K}\|\mathbf{w}_k\|_2^2 \le P_{\mathrm{tot}},
\end{aligned}
\end{equation}
where $\mathbf{W}=[\mathbf{w}_1,\ldots,\mathbf{w}_K]\in\mathbb{C}^{M_t\times K}$ and $\gamma_k(\mathbf{W})$ is the user-$k$ SINR defined in \eqref{eq:sinr_define}.

In implementation, the above problem is solved by bisection over the target SINR level $\tau$ together with SOC~\cite{gershman2010convex} feasibility checks.
For a fixed $\tau$, let $s_k=\Re\{\mathbf{h}_k^H\mathbf{w}_k\}$ and impose the phase-fixing constraints
$
\Im\{\mathbf{h}_k^H\mathbf{w}_k\}=0,\ s_k\ge 0,\ \forall k\in\mathcal{K}.
$
Then, with $\sigma^2$ denoting the receiver noise variance defined in the system model, the SINR feasibility condition can be written in SOC form as
\begin{equation}
\left\|
\begin{bmatrix}
\{\Re(\mathbf{h}_k^H\mathbf{w}_j)\}_{j\neq k} \\
\{\Im(\mathbf{h}_k^H\mathbf{w}_j)\}_{j\neq k} \\
\sqrt{\sigma^2}
\end{bmatrix}
\right\|_2
\le
\frac{1}{\sqrt{\tau}}\, s_k,
\qquad \forall k\in\mathcal{K}.
\end{equation}
By performing bisection on $\tau$ and solving the associated second-order-cone programming (SOCP) feasibility problem, we obtain a high-quality beamforming solution denoted by
$
\mathbf{W}_{\mathrm{soc}}=[\mathbf{w}_{1,\mathrm{soc}},\ldots,\mathbf{w}_{K,\mathrm{soc}}]\in\mathbb{C}^{M_t\times K}.
$

We convert $\mathbf{W}_{\mathrm{soc}}$ into the DFT domain target $\mathbf{X}_0^{\mathrm{soc}}\in\mathbb{R}^{2K\times M_t}$ by stacking the real and imaginary parts of the DFT coefficients of $\{\mathbf{w}_{k,\mathrm{soc}}\}_{k=1}^{K}$.
This pseudo-label is then used as the supervision target for teacher-student training.

\subsubsection{Why single target training is insufficient}
Although the SOC solution provides a high-quality supervision target, it is not the only beamforming output that can achieve comparable performance. A simple example is the phase ambiguity of the max--min SINR objective: replacing $\mathbf{w}_k$ by $\mathbf{w}_k e^{j\phi_k}$ does not change $|\mathbf{h}_i^H\mathbf{w}_k|^2$ for any $i,k$, and therefore leaves all users' SINRs unchanged~\cite{candes2013phaselift}. This multi-solution structure is further enlarged by limited RSRP observations, since similar prompts may correspond to different channel conditions, each of which may admit multiple effective sets of joint beamforming weights. Therefore, forcing the diffusion beamformer to match a single SOC target may collapse the generated outputs toward one selected representation. We use the SOC pseudo-label as a strong supervision target rather than the only correct target, with the EMA teacher providing stable guidance while preserving generative flexibility.

\subsubsection{Teacher-online and EMA teacher}
Based on the above discussion, we first train a teacher-online denoiser using the SOC pseudo-label as a strong supervision target. The teacher-online denoiser learns a SOC-guided denoising target that provides structured supervision for the student. Let $\psi$ denote the parameters of the teacher-online denoiser.
Using the SOC-derived target representation $\mathbf{X}_0^{\mathrm{soc}}$, the teacher-online network is trained under the forward diffusion process as
\begin{equation}
L_T
=
\mathbb{E}_{\mathbf{X}_0^{\mathrm{soc}},t,\boldsymbol{\epsilon}}
\!\left[
\left\|
\boldsymbol{\epsilon}
-
\epsilon_{\psi}(\mathbf{X}_t,t,\mathbf{O})
\right\|_F^2
\right],
\end{equation}
where $\mathbf{X}_t$ is generated by corrupting $\mathbf{X}_0^{\mathrm{soc}}$ according to the forward diffusion process defined in \eqref{eq:forward_diffusion_reparam}.
In this way, the teacher-online model learns to denoise toward SOC-guided beam representations under full prompt conditioning.

To improve target stability during distillation, we further maintain an exponential moving average (EMA)~\cite{hinton2015distilling} copy of the teacher-online network.
Let $\bar{\psi}$ denote the EMA teacher parameters.
After each teacher-online update, the EMA teacher is updated as following
\begin{equation}
\bar{\psi}
\leftarrow
\mu \bar{\psi}
+
(1-\mu)\psi,
\end{equation}
where $\mu\in(0,1)$ is the EMA decay factor.
Compared with the rapidly changing online teacher, the EMA teacher provides a temporally smoothed supervision signal and serves as a more stable distillation target for the student model.

\subsubsection{Student objectives}
Let $\theta$ denote the student network parameters, and define $\tilde{\mathbf{O}}$ as the masked prompt used by the student.
The student is trained under three complementary objectives, namely, a diffusion space denoising loss, a teacher-guided distillation loss, and a task related SINR-oriented guidance loss.

\emph{First}, the student is required to predict the injected Gaussian noise directly from the corrupted sample, leading to the denoising objective
\begin{equation}
L_{\epsilon}
=
\mathbb{E}_{\mathbf{X}_0^{\mathrm{soc}},t,\boldsymbol{\epsilon}}
\!\left[
\left\|
\boldsymbol{\epsilon}
-
\epsilon_{\theta}(\mathbf{X}_t,t,\tilde{\mathbf{O}})
\right\|_F^2
\right].
\end{equation}
This term preserves the basic diffusion denoising capability of the student and ensures consistency with the target reconstruction process.

\emph{Second}, to distill the knowledge of the EMA teacher, the student is encouraged to match the teacher's predicted noise under the same corrupted sample:
\begin{equation}
L_{\mathrm{KD}}
=
\mathbb{E}_{\mathbf{X}_0^{\mathrm{soc}},t,\boldsymbol{\epsilon}}
\!\left[
\left\|
\epsilon_{\theta}(\mathbf{X}_t,t,\tilde{\mathbf{O}})
-
\epsilon_{\bar{\psi}}(\mathbf{X}_t,t,\mathbf{O})
\right\|_F^2
\right].
\end{equation}
Here, the EMA teacher receives the full prompt $\mathbf{O}$, whereas the student uses the masked prompt $\tilde{\mathbf{O}}$.
This design allows the teacher to provide a stable high-quality denoising target, while improving the robustness of the student to imperfect and incomplete prompt information.

\emph{Third}, to inject task-level fairness guidance, we reconstruct the target beam estimate from the student predicted noise as
\begin{equation}
\hat{\mathbf{X}}_0
=
\frac{\mathbf{X}_t-\sqrt{1-\bar{\alpha}_t}\,
\epsilon_{\theta}(\mathbf{X}_t,t,\tilde{\mathbf{O}})}
{\sqrt{\bar{\alpha}_t}},
\end{equation}
and decode it to the beamforming matrix $\hat{\mathbf{W}}$ according to Section~\ref{sec:dit_denoiser}.
Using the resulting SINRs $\{\gamma_k(\hat{\mathbf{W}};\mathbf{H})\}_{k=1}^{K}$, the fairness-oriented guidance loss is defined as
\begin{subequations}\label{eq:sinr_loss}
\begin{align}
&L_{\mathrm{SINR}}
=
-
\mathbb{E}
\!\left[
\mathrm{softmin}_{\tau_s}
\big(
\{10\log_{10}\gamma_k(\hat{\mathbf{W}};\mathbf{H})\}_{k=1}^{K}
\big)
\right],\\
&\mathrm{softmin}_{\tau_s}(\{r_k\}_{k=1}^{K})
=
-\tau_s\log\!\left(\sum_{k=1}^{K}e^{-r_k/\tau_s}\right) \notag\\
&\hphantom{\mathrm{softmin}_{\tau_s}(\{r_k\}_{k=1}^{K})=}
+\tau_s\log K,
\end{align}
\end{subequations}
where $\tau_s>0$ is the soft-min temperature.
This term directly encourages the generated beamforming weights to improve the worst user performance.
The final student loss combines the above objectives as
\begin{equation}
L_S
=
\lambda_{\epsilon}(e)L_{\epsilon}
+
\lambda_{\mathrm{KD}}(e)L_{\mathrm{KD}}
+
\lambda_{\mathrm{SINR}}(e)L_{\mathrm{SINR}},
\label{eq:student_total_loss}
\end{equation}
where $e$ denotes the training epoch, and $\lambda_{\epsilon}(e)$, $\lambda_{\mathrm{KD}}(e)$, and $\lambda_{\mathrm{SINR}}(e)$ are epoch dependent weighting coefficients. We next describe how these loss weights are scheduled together with the diffusion steps used for teacher and student training.

\begin{algorithm}[t]
\caption{Offline KD-EMA Training}
\label{alg:kdema_training}
\footnotesize
\begin{algorithmic}[1]
\REQUIRE Training set $\mathcal{D}$, student denoiser $\epsilon_\theta$, teacher-online denoiser $\epsilon_\psi$, EMA teacher $\epsilon_{\bar{\psi}}$, noise schedule $\{\bar{\alpha}_t\}_{t=1}^{T}$
\ENSURE Trained student denoiser $\epsilon_\theta$

\STATE Initialize $\theta$ and $\psi$, and set $\bar{\psi}\leftarrow\psi$
\FOR{each training epoch}
    \FOR{each mini-batch $(\mathbf{O},\mathcal{H})\sim\mathcal{D}$}
        \STATE Compute SOC pseudo-labels $\mathbf{W}_{\mathrm{soc}}$ from $\mathcal{H}$
        \STATE Convert $\mathbf{W}_{\mathrm{soc}}$ into the target $\mathbf{X}_0^{\mathrm{soc}}$
        \STATE Sample timestep $t$ and noise $\boldsymbol{\epsilon}\sim\mathcal{N}(\mathbf{0},\mathbf{I})$
        \STATE Construct $\mathbf{X}_t$ using \eqref{eq:forward_diffusion_reparam}
        \STATE Update $\epsilon_\psi$ with full prompt $\mathbf{O}$ using the teacher denoising loss
        \STATE Update EMA teacher: $\bar{\psi}\leftarrow \mu\bar{\psi}+(1-\mu)\psi$
        \STATE Generate masked prompt $\widetilde{\mathbf{O}}$ for student training
        \STATE Obtain detached EMA guidance from $\epsilon_{\bar{\psi}}(\mathbf{X}_t,t,\mathbf{O})$
        \STATE Update $\epsilon_\theta$ using the scheduled student loss in \eqref{eq:student_total_loss}
    \ENDFOR
\ENDFOR
\STATE \textbf{return} $\epsilon_\theta$
\end{algorithmic}
\end{algorithm}

\subsubsection{Optimization schedule}
The optimization schedule contains two coupled parts: diffusion step sampling for the teacher and student, and epoch related weighting of the student losses.

The first part is motivated by the noise related behavior of diffusion denoising.
At larger diffusion steps, the sample is heavily corrupted, and errors are more strongly amplified in the reconstructed target estimate according to \eqref{eq:x0_error_relation}.
Consistent with observations in diffusion models, high noise denoising is more associated with coarse structure, whereas low noise denoising is more related to local refinement around a possible target solution~\cite{stracke2025cleandift,pavlova2025diffusion}.

This motivates using different denoising roles for the teacher and student instead of training a single DiT branch over the entire reverse process.
This separation also follows the broader view that diffusion training and sampling choices can be designed separately across noise regimes~\cite{karras2022elucidating}.
The SOC solution provides a high-quality but selected supervision target.
If one branch is trained for both high noise recovery and low noise refinement toward this target, the reverse trajectory can be overly tied to the selected representative solution.
In our framework, the teacher-online branch is trained more often at higher noise steps to learn coarse beamforming structures from the SOC supervision. The EMA teacher provides a stable distillation target, while the masked prompt student focuses more on lower noise denoising to learn general refinement patterns for beamforming weights under imperfect observations. This design guides the student with SOC supervision without forcing it to match a single SOC solution.

The second part controls the weights of the three student losses in \eqref{eq:student_total_loss}. At the beginning of training, the student is mainly trained by $L_{\epsilon}$ and $L_{\mathrm{KD}}$, while $L_{\mathrm{SINR}}$ is disabled. This avoids unstable SINR gradients before the generated beamforming weights have learned meaningful structures. As training proceeds, the weights of $L_{\epsilon}$ and $L_{\mathrm{KD}}$ are reduced, and the weight of $L_{\mathrm{SINR}}$ is increased. In this way, the student first learns stable denoising behavior from the SOC guided teacher, and then gradually shifts toward improving the worst user SINR. This schedule is also consistent with the observation that guidance in diffusion models can depend on the noise interval and may not be equally useful at all diffusion steps~\cite{kynkaanniemi2024applying}. The complete offline training procedure is summarized in Algorithm~\ref{alg:kdema_training}.

\subsection{Inference with DDIM and Candidate Selection}
\label{sec:ddim_infer}

\subsubsection{Student-only deployment}
During online deployment, only the trained student denoiser is deployed; the teacher-online network, EMA teacher, and SOC solver are not used in online deployment.
Given the channel probing based prompt $\mathbf{O}$, the student constructs the conditioning representation in Section~\ref{sec:io_cond} and performs conditional reverse denoising from Gaussian noise.
Thus, online beam generation depends only on the low-dimensional RSRP observations and the student model, without requiring CSI at the BS or online SOC optimization.

\subsubsection{DDIM sampling and candidate generation}
To reduce inference latency, we adopt DDIM sampling~\cite{song2021denoising} instead of the original stochastic denoising diffusion probabilistic models (DDPM) reverse process.
Compared with DDPM, DDIM can generate high-quality solutions with substantially fewer reverse steps, and is therefore more suitable for online beamforming deployment under latency constraints.

Let $\mathcal{T}_{\mathrm{inf}}=\{t_M,\ldots,t_0\}$ denote the DDIM reverse schedule with $t_0=0$ and $t_M=T$.
Starting from an initial Gaussian sample $\mathbf{X}_{t_M}^{(c)} \sim \mathcal{N}(\mathbf{0},\mathbf{I})$, the student denoiser iteratively predicts the noise tensor
\begin{equation}
\hat{\boldsymbol{\epsilon}}_{m}^{(c)}
=
\epsilon_{\theta}\!\left(\mathbf{X}_{t_m}^{(c)},\, t_m,\, \mathbf{O}\right),
\end{equation}
for candidate index $c$ at reverse step $t_m$.
At reverse step $t_m$, the corresponding target estimate is computed analogously to \eqref{eq:x0_reconstruction}, and is denoted by $\hat{\mathbf{X}}_{0,m}^{(c)}$.
The DDIM update is then written as
\begin{equation}
\begin{aligned}
\mathbf{X}_{t_{m-1}}^{(c)}
=
\sqrt{\bar{\alpha}_{t_{m-1}}}\,\hat{\mathbf{X}}_{0,m}^{(c)}
&+
\sqrt{1-\bar{\alpha}_{t_{m-1}}-\tilde{\sigma}_m^{2}}\,
\hat{\boldsymbol{\epsilon}}_{m}^{(c)} \\
&\quad +
\tilde{\sigma}_m \mathbf{z}_{m}^{(c)} ,
\end{aligned}
\label{eq:ddim_update}
\end{equation}
where $\mathbf{z}_{m}^{(c)}\sim\mathcal{N}(\mathbf{0},\mathbf{I})$ if $\eta_{\mathrm{ddim}}>0$, and $\mathbf{z}_{m}^{(c)}=\mathbf{0}$ for deterministic DDIM.
The stochasticity coefficient is
\begin{equation}
\tilde{\sigma}_m^{2}
=
\eta_{\mathrm{ddim}}^{2}
\frac{1-\bar{\alpha}_{t_{m-1}}}{1-\bar{\alpha}_{t_m}}
\left(
1-\frac{\bar{\alpha}_{t_m}}{\bar{\alpha}_{t_{m-1}}}
\right).
\end{equation}

After completing the reverse process, we obtain the final target estimate $\hat{\mathbf{X}}_{0}^{(c)}=\mathbf{X}_{t_0}^{(c)}$, which is then decoded into the resulting beamforming matrix. To improve robustness against sampling variability and mode bias, we repeat the above DDIM procedure for $N_{\mathrm{cand}}$ independent initializations, thereby obtaining the candidate set
$
\mathcal{W}_{\mathrm{cand}}
=
\left\{
\hat{\mathbf{W}}^{(1)},\ldots,\hat{\mathbf{W}}^{(N_{\mathrm{cand}})}
\right\}.
$

\subsubsection{Candidate selection}
Among the generated candidates, we select the final beamforming solution according to a feedback based utility ranking rule.
Specifically,
\begin{equation}
\hat{\mathbf{W}}^{\star}
=
\arg\max_{\hat{\mathbf{W}}^{(c)}\in\mathcal{W}_{\mathrm{cand}}}
U\!\left(\hat{\mathbf{W}}^{(c)}\right),
\end{equation}
where $U(\cdot)$ denotes the candidate selection utility.

% To evaluate each candidate in a deployable manner, the BS transmits short pilot signals using the beamforming matrix $\hat{\mathbf{W}}^{(c)}$.
% For user $k$, the resulting desired-signal and aggregate interference powers under candidate $\hat{\mathbf{W}}^{(c)}$ are given by
% $
% S_k^{(c)} = \left|\mathbf{h}_k^H \hat{\mathbf{w}}_k^{(c)}\right|^2
% $
% and
% $
% I_k^{(c)} = \sum_{i\neq k}\left|\mathbf{h}_k^H \hat{\mathbf{w}}_i^{(c)}\right|^2,
% $
% respectively.
% Operationally, these quantities are measured at the UE side after pilot transmission and then fed back to the BS for candidate reranking.

To evaluate each candidate in a deployable manner, the BS transmits short pilot signals using the beamforming matrix $\hat{\mathbf{W}}^{(c)}$.
For notation, define the desired signal power and interference power of user $k$ under candidate $\hat{\mathbf{W}}^{(c)}$ as
$
S_k^{(c)}=
|\mathbf{h}_k^H\hat{\mathbf{w}}_k^{(c)}|^2
$
and
$
I_k^{(c)}=
\sum_{i\ne k}
|\mathbf{h}_k^H\hat{\mathbf{w}}_i^{(c)}|^2
$.
During online candidate evaluation, user $k$ estimates these two quantities from the pilot measurements and feeds back the scalar estimates $\hat{S}_k^{(c)}$ and $\hat{I}_k^{(c)}$.
Using the scalar estimates fed back by the UEs, the BS assigns each candidate the following utility:
\begin{equation}
U\!\left(\hat{\mathbf{W}}^{(c)}\right)
=
\min_{k\in\mathcal{K}}
\frac{\hat{S}_k^{(c)}}{\hat{I}_k^{(c)}}.
\label{eq:feedback_candidate_utility}
\end{equation}
The final beamforming weights are selected as the candidate with the largest utility.
This selection uses only scalar UE feedback and does not require instantaneous CSI acquisition at the BS.
Since this work focuses on the multi-user setting dominated by interference, the candidate utility does not explicitly include the noise term. Instead, the generated candidates are ranked by the measured desired signal to interference balance, which provides a lightweight selection rule without noise power estimation.
The complete deployable inference and candidate selection procedure is summarized in Algorithm~\ref{alg:ddim_candidate_infer}.

\begin{algorithm}[t]
\caption{Deployable DDIM Inference}
\label{alg:ddim_candidate_infer}
\footnotesize
\begin{algorithmic}[1]
\REQUIRE $\mathbf{O}$, $\epsilon_\theta$, $\mathcal{T}_{\mathrm{inf}}$, $N_{\mathrm{cand}}$, $\eta_{\mathrm{ddim}}$
\ENSURE $\hat{\mathbf{W}}^\star$

\FOR{$c=1$ to $N_{\mathrm{cand}}$}
    \STATE Sample $\mathbf{X}_{t_M}^{(c)}\sim\mathcal{N}(\mathbf{0},\mathbf{I})$
    \FOR{$m=M,\ldots,1$}
        \STATE Predict $\hat{\boldsymbol{\epsilon}}_m^{(c)}
        \leftarrow \epsilon_\theta(\mathbf{X}_{t_m}^{(c)},t_m,\mathbf{O})$
        \STATE Update $\mathbf{X}_{t_{m-1}}^{(c)}$ by the DDIM step in \eqref{eq:ddim_update}
    \ENDFOR
    \STATE Decode $\mathbf{X}_{t_0}^{(c)}$ into $\hat{\mathbf{W}}^{(c)}$
\ENDFOR

\IF{$N_{\mathrm{cand}}=1$}
    \STATE \textbf{return} $\hat{\mathbf{W}}^{(1)}$
\ENDIF

\FOR{$c=1$ to $N_{\mathrm{cand}}$}
    \STATE Transmit short pilots using $\hat{\mathbf{W}}^{(c)}$
    \STATE Each UE estimates $\hat{S}_k^{(c)}$ and $\hat{I}_k^{(c)}$ and feeds them back to the BS
    \STATE Compute
    $
    U(\hat{\mathbf{W}}^{(c)})
    \leftarrow
    \min_{k\in\mathcal{K}}
    \hat{S}_k^{(c)}/\hat{I}_k^{(c)}
    $
\ENDFOR

\STATE $\hat{\mathbf{W}}^\star
\leftarrow
\arg\max_{\hat{\mathbf{W}}^{(c)}} U(\hat{\mathbf{W}}^{(c)})$
\STATE \textbf{return} $\hat{\mathbf{W}}^\star$
\end{algorithmic}
\end{algorithm}

\section{Comparison with Baseline Approaches}
\label{sec:baseline_compare}

We compare KDGen-BF with conventional codebook-based baselines, a site-specific learned-codebook baseline, and site-specific grid-free learning baselines under a unified online protocol.
All deployable methods use low-dimensional RSRP observations and do not assume instantaneous CSI at the BS during online inference.
Whenever a method produces multiple candidate beams or beam combinations, the final solution is selected using the same feedback based utility in \eqref{eq:feedback_candidate_utility}.
This shared selection rule accounts for interference during candidate selection and ensures a fair comparison among the methods in multi-user settings.

\subsection{Codebook-Based Baselines}
\label{sec:codebook_baselines}

We first consider codebook-based baselines whose final beams are selected from finite DFT codebooks.
Let $\mathcal{F}_{\mathrm{p}}=\{\mathbf{f}_1,\ldots,\mathbf{f}_{N_p}\}$ denote the probed DFT beam subset.
For a selected beam index combination $\mathbf{m}=[m_1,\ldots,m_K]^T$, the beamforming vector of user $k$ is
\begin{equation}
\mathbf{w}_k(\mathbf{m})
=
\sqrt{\frac{P_{\mathrm{tot}}}{K}}\mathbf{f}_{m_k},
\qquad k\in\mathcal{K}.
\end{equation}

\subsubsection{DFT Greedy}
DFT Greedy selects each user's beam independently according to the largest probing RSRP,
\begin{equation}
m_k^{\mathrm{greedy}}
=
\arg\max_{n\in\{1,\ldots,N_p\}}
p_{k,n}.
\end{equation}
The resulting combination is directly used without joint candidate selection, and therefore does not explicitly account for inter-user interference.

\subsubsection{DFT Exhaustive}
DFT Exhaustive performs joint combination selection over all probed DFT beams.
Specifically, it enumerates $\mathcal{M}_{\mathrm{exh}}=\{1,\ldots,N_p\}^{K}$, maps each combination $\mathbf{m}\in\mathcal{M}_{\mathrm{exh}}$ to $\mathbf{W}(\mathbf{m})$, evaluates it using the feedback based utility in \eqref{eq:feedback_candidate_utility}, and selects the best combination.
This baseline accounts for inter-user coupling but requires combinatorial combination enumeration.

\subsubsection{Hier-DFT}
Hier-DFT reduces the exhaustive combination space through coarse-to-fine DFT probing.
The full DFT codebook is partitioned into $N_c$ angular sectors, and one representative beam from each sector is used for coarse probing.
According to the coarse RSRP feedback, each user keeps its top-$B_c$ sectors.
Fine probing is then performed over $N_f$ DFT beams selected from these kept sectors.
After fine probing, user $k$ keeps its top-$B$ fine beams to form $\mathcal{M}_k^{\mathrm{hier}}$.
The BS then evaluates the reduced combination set
$
\mathcal{M}_{\mathrm{hier}}
=
\mathcal{M}_1^{\mathrm{hier}}\times\cdots\times\mathcal{M}_K^{\mathrm{hier}},
$
which contains $B^K$ candidate combinations, using \eqref{eq:feedback_candidate_utility} and selects the best combination.

\subsection{Learned-Codebook Baseline}
\label{sec:learned_codebook_baseline}

\subsubsection{NN-LSS-Codebook}
We further include NN-LSS-Codebook, where NN-LSS denotes neural-network learned site-specific, as a learned site-specific codebook baseline.
Unlike DFT-based methods that use fixed angular beams, this baseline trains a site-specific beam codebook offline and performs online combination selection over learned codewords.

Let $\mathcal{G}=\{\mathbf{g}_1,\ldots,\mathbf{g}_{M}\}$ denote the learned codebook, where each codeword $\mathbf{g}_m\in\mathbb{C}^{M_t\times 1}$ is normalized.
The codebook is trained from the training split using SOC-derived beamforming targets.
During online evaluation, each user reports a score $q_{k,m}^{\mathrm{nn}}$ for codeword $\mathbf{g}_m$ through UE-side beam feedbacks.
For each user, the top-$B$ codewords are kept to form the candidate set $\mathcal{M}_k^{\mathrm{nn}}\subseteq\{1,\ldots,M\}$.
The BS then evaluates the reduced combination set
$
\mathcal{M}_{\mathrm{nn}}
=
\mathcal{M}_1^{\mathrm{nn}}\times\cdots\times\mathcal{M}_K^{\mathrm{nn}}
$
using \eqref{eq:feedback_candidate_utility} and selects the best combination.

\subsection{Grid-Free Learning-Based Baselines}
\label{sec:grid_free_learning_baselines}

We next consider site-specific grid-free learning baselines, whose outputs are not restricted to a finite beam codebook.

\subsubsection{Deterministic Transformer}
This baseline uses the same group prompt $\mathbf{O}$ as KDGen-BF, but directly predicts the target beam representation without diffusion sampling:
\begin{equation}
\hat{\mathbf{X}}_{0}^{\mathrm{tr}}
=
f_{\mathrm{tr}}(\mathbf{O}).
\end{equation}
The output $\hat{\mathbf{X}}_{0}^{\mathrm{tr}}$ is decoded into $\hat{\mathbf{W}}_{\mathrm{tr}}$ using the same DFT domain decoding, inverse Fourier transform, and total power normalization.
It therefore serves as a deterministic grid-free counterpart to the proposed generative beamformer.

\begin{table*}[t]
\centering
\caption{Dataset, training, inference, and baseline parameters.}
\label{tab:simulation_parameters}
\scriptsize
\setlength{\tabcolsep}{5pt}
\renewcommand{\arraystretch}{1.08}
\begin{tabular}{l c l c}
\hline
\textbf{Parameter} & \textbf{Value} &
\textbf{Parameter} & \textbf{Value} \\
\hline
\multicolumn{4}{l}{\textit{Dataset setup}} \\
Scenarios & \texttt{I2\_28B}, \texttt{O1B\_28}, \texttt{Boston5G\_28}
& Carrier frequency & 28 GHz \\
BS array & 64-element ULA
& Antenna spacing & $0.5\lambda$ \\
UE antenna & Single antenna
& Users & $K\in\{2,3,4\}$ \\
Groups & 102,400 per scenario
& Split & $8{:}1{:}1$ \\
Probed DFT beams & $N_p=64$ default
& RSRP noise setting & SNR = 40 dB \\
\hline
\multicolumn{4}{l}{\textit{Training setup}} \\
Optimizer & AdamW
& Learning rate & $10^{-4}$ \\
Batch size & 64
& Epochs & 1000 \\
Diffusion steps & 1000
& EMA decay & 0.995 \\
\hline
\multicolumn{4}{l}{\textit{Inference setup}} \\
DDIM steps & $N_{\mathrm{step}}=50$
& DDIM stochasticity & $\eta_{\mathrm{ddim}}=0$ \\
Default candidates & $N_{\mathrm{cand}}=64$
& Candidate sweep & $N_{\mathrm{cand}}\in\{1,2,4,8,16,24,32,64\}$ \\
\hline
\multicolumn{4}{l}{\textit{Baseline setup}} \\
Candidate number & $B=4$
&  Candidate search budget & $B^K$\\
Learned codebook size  & 64
& SU candidates & $M=64$ \\[1pt]
\hline
\end{tabular}
\end{table*}

\subsubsection{Single-user Direction-SOC}
This baseline adapts single-user generative site-specific beamforming to the multi-user setting.
For each user $k$, a single-user generator produces $N_{\mathrm{cand}}^{\mathrm{su}}$ candidate transmit beam directions from prompt $\mathbf{o}_k$:
$
\mathcal{V}_k^{\mathrm{su}}
=
\{\tilde{\mathbf{v}}_{k}^{(1)},\ldots,
\tilde{\mathbf{v}}_{k}^{(N_{\mathrm{cand}}^{\mathrm{su}})}\},
\quad
\tilde{\mathbf{v}}_{k}^{(c)}\in\mathbb{C}^{M_t\times 1}.
$
The candidate with the largest UE-side received power is selected
\begin{equation}
\tilde{\mathbf{v}}_k^{\star}
=
\arg\max_{\tilde{\mathbf{v}}\in\mathcal{V}_k^{\mathrm{su}}}
q_k(\tilde{\mathbf{v}}),
\end{equation}
where $q_k(\tilde{\mathbf{v}})$ is the RSRP measured by user $k$ under transmit beam $\tilde{\mathbf{v}}$.
Since this single-user selection is MRT-like, the normalized beam
$
\hat{\mathbf{v}}_k=
\tilde{\mathbf{v}}_k^{\star}/\|\tilde{\mathbf{v}}_k^{\star}\|_2
$
is used as an estimated channel direction.
With
$
\hat{\mathbf{H}}_{\mathrm{su}}
=
[\hat{\mathbf{v}}_1,\ldots,\hat{\mathbf{v}}_K],
$
we solve the SOC-based max--min beamforming problem using $\hat{\mathbf{H}}_{\mathrm{su}}$, yielding
$
\hat{\mathbf{W}}_{\mathrm{su+soc}}
=
\mathrm{SOC}(\hat{\mathbf{H}}_{\mathrm{su}}).
$

\begin{table}[t]
\caption{BS-side serial decision complexity for one $K$-user group.}
\label{tab:complexity_comparison}
\centering
\small
\setlength{\tabcolsep}{4pt}
\begin{tabular}{@{}l l@{}}
\hline
\textbf{Method} & \textbf{Complexity} \\
\hline
DFT Greedy & $\mathcal{O}(K N_p)$ \\
DFT Exhaustive & $\mathcal{O}(N_p^K)$ \\
Hier-DFT & $\mathcal{O}(K(N_c+N_f)+B^K)$ \\
NN-LSS-Codebook & $\mathcal{O}(K M+B^K)$ \\
Det. Transformer & $\mathcal{O}(C_{\mathrm{NN}})$ \\
SU Dir.-SOC & $\mathcal{O}(K N_{\mathrm{cand}}^{\mathrm{su}} N_{\mathrm{step}} C_{\mathrm{NN}} + C_{\mathrm{soc}})$ \\
KDGen-BF & $\mathcal{O}(N_{\mathrm{cand}} N_{\mathrm{step}} C_{\mathrm{NN}})$ \\
\hline
\end{tabular}
\end{table}

\subsection{Complexity Analysis}
\label{sec:decision_complexity}

We then compare the BS side serial decision complexity for one $K$ user group.
The decision time is measured after the required probing observations or UE reported scores are available and before downlink data transmission.
Over the air probing, UE side measurement, and feedback signaling are excluded, while BS side RSRP processing and joint candidate selection are counted.
For a fair comparison and to avoid weakening the baselines, the feedback based joint candidate search baselines are allowed to keep $B$ candidate beams for each user and evaluate all $B^K$ joint beam combinations.
Thus, their over the air candidate evaluation overhead scales with $B^K$, whereas the corresponding overhead of KDGen-BF is controlled by the number of generated candidates $N_{\mathrm{cand}}$.

Let $C_{\mathrm{NN}}$ denote the architecture dependent cost of one neural network forward pass, and let $C_{\mathrm{soc}}$ denote the cost of one SOC based max--min beamforming solve. We use a single $C_{\mathrm{NN}}$ notation for compactness. For multi-user neural methods, one forward pass processes one whole $K$ user group, so the dependence on $K$ is included in $C_{\mathrm{NN}}$ through the input and output dimensions. For single-user diffusion generators, the denoiser is applied separately to each user, each candidate, and each DDIM reverse step, resulting in the factor $K N_{\mathrm{cand}}^{\mathrm{su}} N_{\mathrm{step}}$. For feedback based joint candidate search baselines, $B$ denotes the kept candidate size for each user, yielding $B^K$ joint candidate combinations. For Hier-DFT, $N_c$ and $N_f$ denote the numbers of coarse and fine probing beams, respectively. For NN-LSS-Codebook, $M$ denotes the learned codebook size. The complexity of each method is summarized in Table~\ref{tab:complexity_comparison}, where lower order utility aggregation costs are omitted for compactness.

As shown in Table~\ref{tab:complexity_comparison}, only DFT Exhaustive has inherent exponential scaling over the full probed codebook, i.e., $N_p^K$.
For Hier-DFT and NN-LSS-Codebook, the $B^K$ term is an optional feedback based cost used to strengthen multi-user selection; without it, their main scoring costs scale linearly with $K$.
KDGen-BF requires $N_{\mathrm{cand}}N_{\mathrm{step}}$ multi-user student denoiser evaluations, with all teacher and SOC pseudo-label operations kept offline.

\section{Numerical Results}
\label{sec:results}

\subsection{Experimental Setup}
\label{sec:exp_setup}

We evaluate the proposed method on three DeepMIMO~\cite{alkhateeb2019deepmimo} ray-tracing scenarios: \texttt{I2\_28B}, \texttt{O1B\_28}, and \texttt{Boston5G\_28}, representing indoor, outdoor street-level, and dense urban propagation environments, respectively.
For each scenario, we construct multi-user groups with $K\in\{2,3,4\}$ under the downlink MISO setting in Section~\ref{sec:system_model}.
The same preprocessing, grouping, probing, and data-splitting protocol is used for all compared methods.

All models are implemented in PyTorch and trained on a server with NVIDIA A100-SXM4-80GB GPUs.
A separate model is trained for each scenario and each value of $K$.
Unless otherwise specified, the default probing budget is $N_p=64$, the DDIM reverse-step number is $N_{\mathrm{step}}=50$, and the default candidate number is $N_{\mathrm{cand}}=64$.
The RSRP noise setting in Table~\ref{tab:simulation_parameters} is used to corrupt probing observations, while all reported performance metrics are based on downlink SINR and include inter-user interference.

For joint candidate search baselines, the kept candidate size is $B=4$ unless otherwise specified.
For Hier-DFT, the coarse and fine split is budget-matched to $N_p$, with $(N_c,N_f)=(2,2),(4,4),(4,12),(8,24),(8,40),(8,56)$ for $N_p=4,8,16,32,48,64$, respectively.
The kept coarse-sector number is $B_c=1$ for $N_p=4$ and $B_c=2$ otherwise, and the final kept fine-beam number is $B=2$ for $N_p=4$ and $B=4$ otherwise.
The detailed dataset, training, inference, and baseline configurations are summarized in Table~\ref{tab:simulation_parameters}.

\subsection{Evaluation under Different Probing Budgets}
\label{sec:exp_main_results}

\begin{figure*}[!t]
    \centering
    \includegraphics[width=1.0\textwidth]{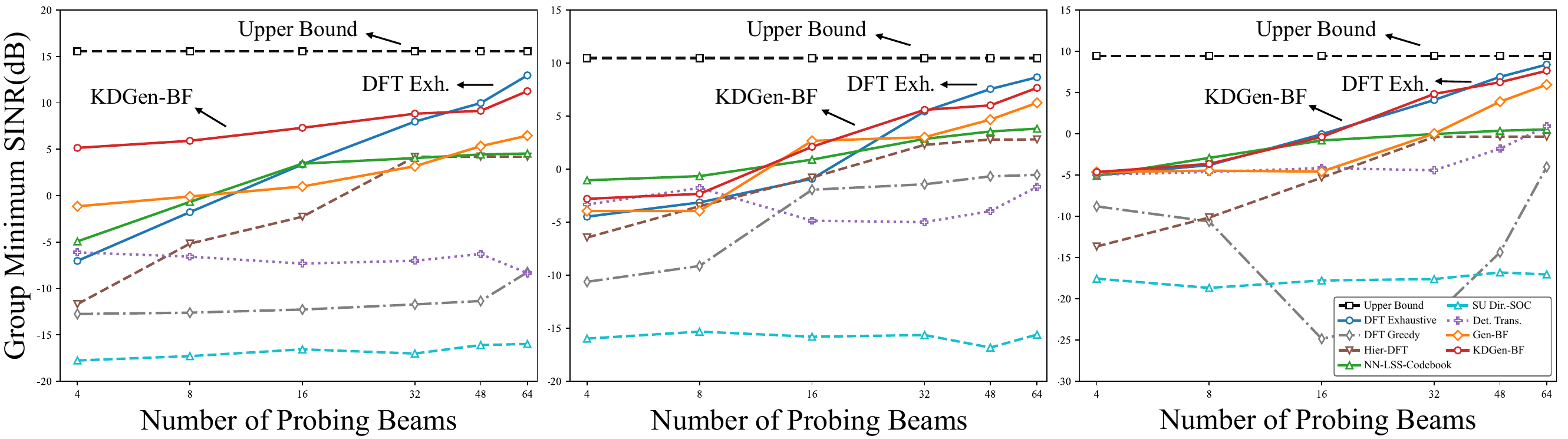}
    \caption{Mean group minimum SINR versus the number of probed DFT beams $N_p$ for $K=4$ users under three DeepMIMO scenarios.}
    \label{fig:main_np}
\end{figure*}

Fig.~\ref{fig:main_np} reports the mean group minimum SINR under different probing budgets for $K=4$ users in the three DeepMIMO scenarios.
The full CSI upper bound is computed from offline CSI and is therefore independent of $N_p$.

Overall, KDGen-BF achieves the best performance among the non-exhaustive deployable methods in \texttt{Boston5G\_28} over the entire probing range, and remains highly competitive in \texttt{I2\_28B} and \texttt{O1B\_28}.
In \texttt{Boston5G\_28}, KDGen-BF improves from $5.15$ dB at $N_p=4$ to $11.26$ dB at $N_p=64$, consistently outperforming DFT Greedy, Hier-DFT, NN-LSS-Codebook, deterministic Transformer, SU Dir.-SOC, and Gen-BF.
In \texttt{O1B\_28}, KDGen-BF increases from $-4.64$ dB to $7.64$ dB as $N_p$ grows from $4$ to $64$, and becomes the best non-exhaustive method for $N_p\geq 8$.
In \texttt{I2\_28B}, KDGen-BF reaches $7.66$ dB at $N_p=64$, approaching DFT Exhaustive at $8.66$ dB while clearly outperforming the other non-exhaustive baselines.

KDGen-BF is particularly effective under limited or moderate probing budgets.
For example, in \texttt{Boston5G\_28} at $N_p=16$, KDGen-BF achieves $7.31$ dB, compared with $3.37$ dB for DFT Exhaustive, $3.45$ dB for NN-LSS-Codebook, and $0.99$ dB for Gen-BF.
At $N_p=32$, KDGen-BF still surpasses DFT Exhaustive in \texttt{Boston5G\_28} and slightly exceeds it in \texttt{I2\_28B}.
These results indicate that our conditional generation can outperform finite-codebook search when the probed codebook is in limited sized.

When $N_p$ becomes large, DFT Exhaustive becomes more competitive because the enlarged DFT codebook provides finer angular resolution and a larger joint candidate search space.
Nevertheless, at $N_p=64$, the gaps between KDGen-BF and DFT Exhaustive are only $1.70$ dB, $1.00$ dB, and $0.75$ dB on \texttt{Boston5G\_28}, \texttt{I2\_28B}, and \texttt{O1B\_28}, respectively.
Thus, KDGen-BF provides a favorable performance and search-complexity tradeoff, especially when the probing budget is limited or exhaustive combination search is undesirable.

\newcommand{\ubgap}[2]{#1$^{\scriptscriptstyle(-#2)}$}

\begin{table*}[t]
\centering
\caption{Mean group min-SINR (dB) comparison on \texttt{Boston5G\_28}. Superscripts indicate the gap to Upper Bound.}
\label{tab:boston_main_results}
\footnotesize
\renewcommand{\arraystretch}{1.18}

\setlength{\tabcolsep}{3.5pt}
\begin{tabular*}{0.85\textwidth}{@{\extracolsep{\fill}} c c c c c c c c @{}}
\hline
\textbf{Users}
& \textbf{Upper Bound}
& \textbf{DFT Greedy}
& \textbf{DFT Exh.}
& \textbf{Hier-DFT}
& \textbf{NN-LSS-Codebook}
& \textbf{Det. Trans.}
& \textbf{SU Dir.-SOC} \\
\hline
$K=2$
& 35.73
& \ubgap{10.36}{25.37}
& \ubgap{29.70}{6.03}
& \ubgap{22.39}{13.34}
& \ubgap{31.16}{4.57}
& \ubgap{6.75}{28.98}
& \ubgap{-4.67}{40.40} \\
$K=3$
& 20.36
& \ubgap{-1.66}{22.02}
& \ubgap{17.44}{2.92}
& \ubgap{11.02}{9.34}
& \ubgap{16.18}{4.18}
& \ubgap{-0.26}{20.62}
& \ubgap{-12.26}{32.62} \\
$K=4$
& 15.54
& \ubgap{-8.25}{23.79}
& \ubgap{12.96}{2.58}
& \ubgap{4.20}{11.34}
& \ubgap{4.54}{11.00}
& \ubgap{-8.35}{23.89}
& \ubgap{-15.98}{31.52} \\
\hline
\end{tabular*}

\vspace{0.8mm}

\setlength{\tabcolsep}{2.4pt}
\begin{tabular*}{0.85\textwidth}{@{\extracolsep{\fill}} c c c c c c c c c c @{}}
\hline
\textbf{Users}
& \textbf{Method}
& \multicolumn{8}{c}{$N_{\mathrm{cand}}$} \\
\cline{3-10}
&
& \textbf{1}
& \textbf{2}
& \textbf{4}
& \textbf{8}
& \textbf{16}
& \textbf{24}
& \textbf{32}
& \textbf{64} \\
\hline
\multirow{2}{*}{$K=2$}
& Gen-BF
& 13.78 & 16.02 & 17.72 & 19.19 & 20.42 & 21.05 & 21.52 & \ubgap{22.51}{13.22} \\
& KDGen-BF
& 13.70 & 16.11 & 18.06 & 19.65 & 21.05 & 21.77 & 22.26 & \ubgap{27.36}{8.37} \\
\hline
\multirow{2}{*}{$K=3$}
& Gen-BF
& -0.36 & 2.41 & 4.38 & 5.92 & 7.22 & 7.88 & 8.31 & \ubgap{11.29}{9.07} \\
& KDGen-BF
& 1.07 & 4.15 & 6.61 & 8.64 & 10.20 & 10.98 & 11.50 & \ubgap{14.64}{5.72} \\
\hline
\multirow{2}{*}{$K=4$}
& Gen-BF
& -6.26 & -3.39 & -1.08 & 0.71 & 2.15 & 2.92 & 3.43 & \ubgap{6.47}{9.07} \\
& KDGen-BF
& -8.48 & -4.44 & -1.10 & 1.60 & 3.85 & 4.97 & 5.72 & \ubgap{11.26}{4.28} \\
\hline
\end{tabular*}

\end{table*}

\begin{figure}[!t]
    \centering
    \includegraphics[width=0.85\columnwidth]{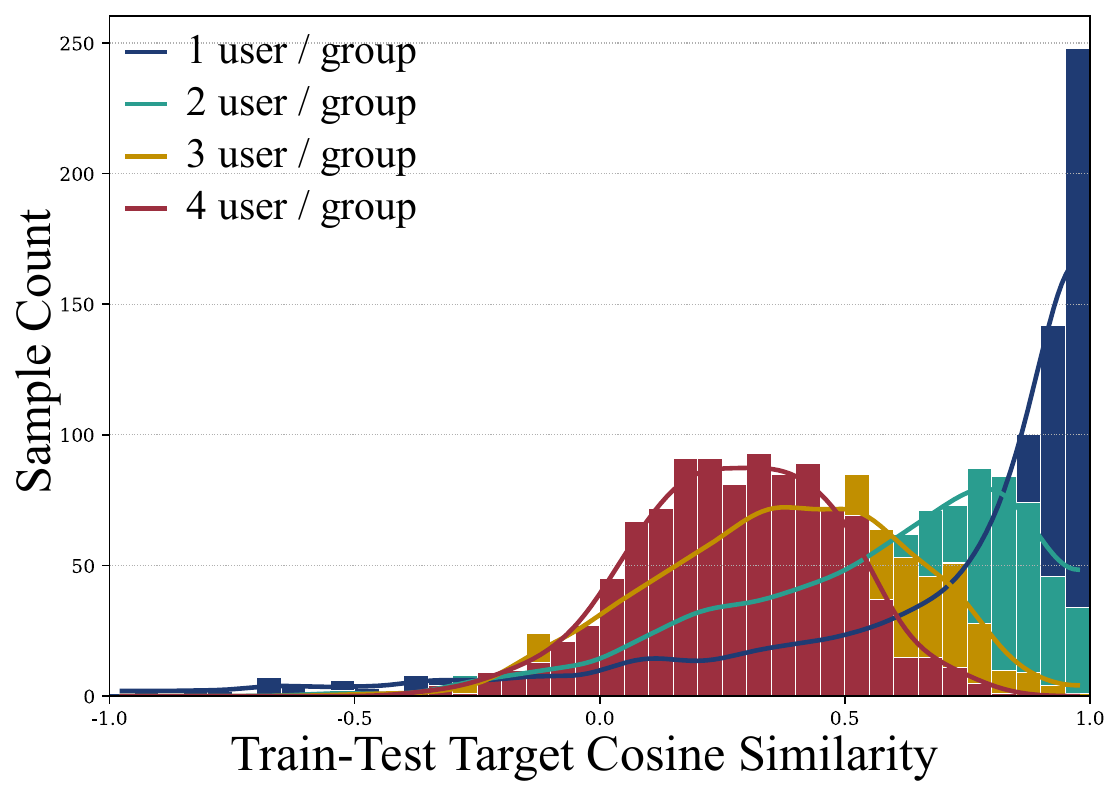}
    \caption{Train-test target cosine similarity distributions on \texttt{Boston5G\_28} under different user-group sizes. For each test target, the similarity is computed with respect to its nearest training target in the target space.}
    \label{fig:target_similarity}
\end{figure}

\subsection{From Single-User Fitting to Multi-User Generalization}
\label{sec:exp_multiuser_generalization}

To explain why multi-user beamforming is not a straightforward extension of single-user beam prediction, we analyze the train-test similarity of target beamforming solutions.
For each test target, we compute its maximum cosine similarity to the training targets after converting the beamforming matrix into the real-valued DFT domain representation used for training.

Fig.~\ref{fig:target_similarity} shows the similarity distributions on \texttt{Boston5G\_28} for different user-group sizes.
For $K=1$, most test targets have high similarity to at least one training target, indicating that the single user target space is well covered by the training set.
As $K$ increases, the distributions shift toward lower values and become more dispersed, which means that multi-user beamforming targets are less likely to have close training-set counterparts.
Thus, multi-user prediction cannot rely only on target space memorization or nearest neighbor interpolation; it must learn joint beamforming structure, including interference coupling and fairness-oriented coordination.

This observation is consistent with the performance trend in Table~\ref{tab:boston_main_results}.
Although the Upper Bound also decreases as $K$ increases due to stronger multi-user interference, the degradation of single user baselines is much more severe.
For example, from $K=2$ to $K=4$, NN-LSS-Codebook drops from $31.16$ dB to $4.54$ dB, the deterministic Transformer drops from $6.75$ dB to $-8.35$ dB, and SU Dir.-SOC drops from $-4.67$ dB to $-15.98$ dB.
Their gaps to the Upper Bound also enlarge substantially, indicating that learned codebook methods, deterministic regression, and single user direction fitting do not adequately capture the coupled tradeoff among desired signal enhancement, inter-user interference suppression, and max--min fairness.
For $K=4$, KDGen-BF reduces the gap to the Upper Bound to $4.28$ dB, compared with $11.00$ dB for NN-LSS-Codebook, $23.89$ dB for the deterministic Transformer, and $31.52$ dB for SU Dir.-SOC.

\subsection{Ablation Studies}
\label{sec:exp_ablation}

Table~\ref{tab:boston_main_results} reports the ablation results on \texttt{Boston5G\_28}.
The lower part of the table compares Gen-BF and KDGen-BF under different candidate numbers, thereby examining the effects of KD-EMA training and feedback based multi-candidate selection.

\subsubsection{Effect of KD-EMA training}
Comparing Gen-BF and KDGen-BF isolates the effect of the proposed KD-EMA teacher-student strategy.
Gen-BF is trained without the SOC-supervised EMA teacher, whereas KDGen-BF uses SOC-guided distillation.
At $N_{\mathrm{cand}}=64$, KDGen-BF consistently outperforms Gen-BF for all user numbers.
The mean group minimum SINR improves from $22.51$ dB to $27.36$ dB for $K=2$, from $11.29$ dB to $14.64$ dB for $K=3$, and from $6.47$ dB to $11.26$ dB for $K=4$.
For the most challenging $K=4$ case, the gap to the full CSI Upper Bound is reduced from $9.07$ dB to $4.28$ dB.

These gains indicate that KD-EMA improves the quality of the generated beamforming distribution, especially when the interference coupled solution space becomes more complex.
The EMA teacher provides a stable SOC-guided denoising target, while the student keeps the flexibility of conditional generation instead of collapsing to a single pseudo-label representative.

\subsubsection{Effect of candidate number}
We next examine the role of the number of generated candidates $N_{\mathrm{cand}}$.
For both Gen-BF and KDGen-BF, a larger $N_{\mathrm{cand}}$ improves the performance by providing more candidate beamforming weights for selection.
For KDGen-BF, as $N_{\mathrm{cand}}$ increases from $1$ to $64$, the mean group minimum SINR improves from $13.70$ dB to $27.36$ dB for $K=2$, from $1.07$ dB to $14.64$ dB for $K=3$, and from $-8.48$ dB to $11.26$ dB for $K=4$.
The gain is most pronounced for $K=4$, indicating that single-sample generation is insufficient in the more strongly coupled multi-user setting.

These results show that multi-candidate generation improves coverage of the conditional beamforming solution space. The final selection uses only scalar desired and interference measurements for each candidate, which avoids instantaneous CSI acquisition at the BS while introducing only lightweight pilot measurement and feedback overhead.
It is worth noting that the candidate budget of KDGen-BF is still modest compared with the feedback based joint candidate search baselines: with the default $B=4$, those baselines evaluate $B^K=64$ combinations for $K=3$ and $B^K=256$ combinations for $K=4$.
Thus, even the largest KDGen-BF setting considered here, $N_{\mathrm{cand}}=64$, does not exceed the combination evaluation budget used by strong feedback based baselines in the $K=4$ case.
Moreover, candidate generation in KDGen-BF is naturally batchable: multiple diffusion trajectories can be sampled in parallel, so the wall-clock latency can be much smaller than the serial scaling with $N_{\mathrm{cand}}$ when sufficient hardware resources are available.
Therefore, $N_{\mathrm{cand}}$ controls a practical tradeoff between beamforming weights quality and inference cost, with serial DDIM complexity growing linearly in the number of generated candidates.

\begin{figure}[!t]
    \centering
    \includegraphics[width=0.8\columnwidth]{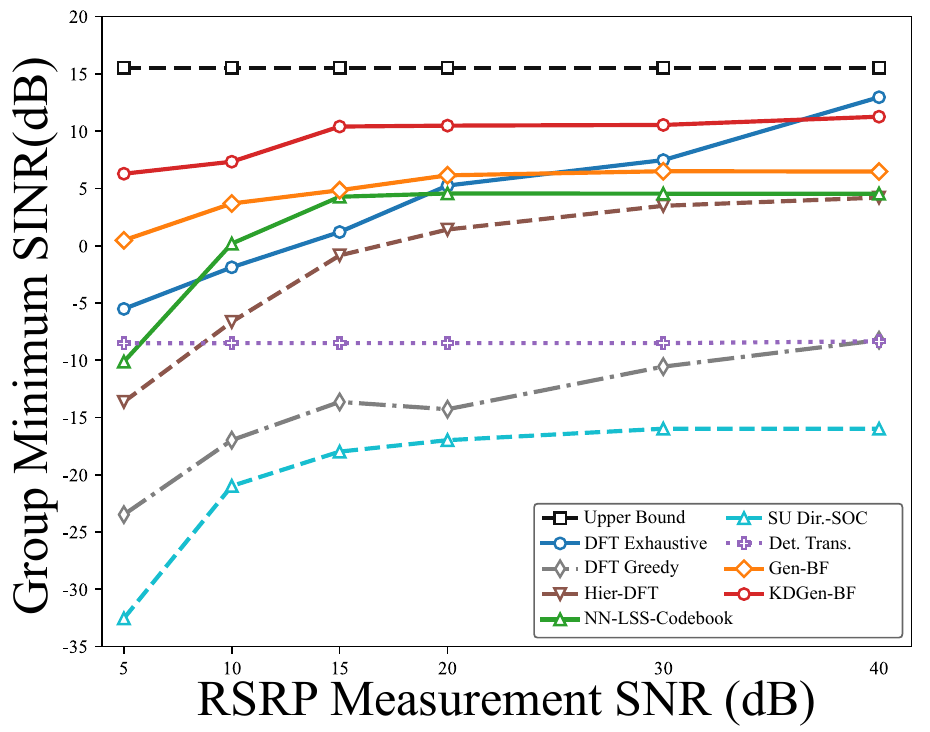}
    \caption{Mean group minimum SINR versus RSRP measurement SNR on \texttt{Boston5G\_28} with $K=4$. Noise is injected into the probing-based RSRP observations to evaluate robustness to prompt corruption.}
    \label{fig:noise_sweep}
\end{figure}

\subsubsection{Robustness to noisy probing observations}
We further evaluate robustness to noisy RSRP measurements.
Different from the receiver noise used in downlink SINR evaluation, the noise here is injected into the low-dimensional RSRP observations before beamforming decisions are made.
Thus, this experiment measures sensitivity to corrupted prompt information rather than robustness to data transmission noise.
The RSRP measurement SNR is swept from $5$ dB to $40$ dB on \texttt{Boston5G\_28} with $K=4$.

Fig.~\ref{fig:noise_sweep} shows that KDGen-BF remains robust even under severely corrupted probing observations.
At an RSRP measurement SNR of $5$ dB, KDGen-BF achieves $6.28$ dB mean group minimum SINR, compared with $0.47$ dB for Gen-BF, $-5.52$ dB for DFT Exhaustive, $-10.13$ dB for NN-LSS-Codebook, and $-13.63$ dB for Hier-DFT.
As the measurement SNR increases to $40$ dB, KDGen-BF reaches $11.26$ dB.
This indicates that the proposed framework is not overly sensitive to RSRP perturbations, even without a dedicated noise robust training module.

The robustness gain is mainly attributed to masked prompt student training and the prompt design.
During training, the student learns from partially masked prompts under the guidance of the full-prompt EMA teacher, which encourages recovery of useful beamforming structure from degraded or incomplete conditioning information.
Meanwhile, the standardized and peak-referenced RSRP features reduce dependence on absolute RSRP scale and emphasize relative beam-domain structure.
Multi-candidate generative inference further improves tolerance to prompt ambiguity, since feedback based strategy selects a high performance beamformer from several plausible candidates.
In contrast, codebook-search and deterministic baselines rely more directly on noisy RSRP rankings or deterministic prompt-to-beam mapping, which makes them more sensitive to prompt corruption.

\section{Conclusion}
\label{sec:conclusion}

% This paper proposed KDGen-BF, a conditional generative framework for CSI-free online multi-user beamforming from low-dimensional RSRP observations.
% KDGen-BF formulates joint beamforming as conditional generation and addresses limited observations, inter-user interference coupling, and multi-solution fairness-oriented beam design through SOC-guided KD-EMA training, masked-prompt learning, and DDIM-based multi-candidate selection.
% Numerical results on three DeepMIMO scenarios show that KDGen-BF achieves strong max--min fairness performance, especially under limited probing budgets and noisy RSRP measurements.
% These results demonstrate the potential of generative modeling for multi-user beamforming when online observations are low-dimensional or imperfect.
% Future work will extend the framework to mobility adaptation, wideband beamforming scenarios, and multi-cell interference coordination.

This paper proposed KDGen-BF, a conditional generative framework for online multi-user beamforming from low-dimensional RSRP observations. The key idea is to replace deterministic beam prediction and finite-codebook selection with joint conditional generation of beamforming weights, so that the model can account for limited observations, interference among users, and the multi-solution nature of fairness-oriented beamforming. Numerical results on multiple DeepMIMO scenarios show that KDGen-BF achieves strong max--min fairness performance, especially under limited probing budgets and noisy RSRP measurements. These results demonstrate the potential of generative modeling for multi-user beamforming when online observations are low-dimensional or imperfect. Future work will extend the framework to mobility adaptation, wideband beamforming scenarios, and multi-cell interference coordination.

\balance
\bibliographystyle{IEEEtran}
\bibliography{reference/mybib}

\end{document}